\documentclass[iop]{emulateapj}
\usepackage{graphicx}
\usepackage{multirow}
\usepackage{float}

\usepackage{color}
\usepackage{amsmath}
\usepackage{ulem}

\usepackage{txfonts} 

\definecolor{brown}{rgb}{0.42,0.24,0.07}
\definecolor{green}{rgb}{0.0,0.6,0.00}
\definecolor{purple}{rgb}{0.7,0.0,0.7}
\definecolor{black}{rgb}{0.0,0.0,0.0}

\newcommand{\beq}{\begin{equation}}
\newcommand{\eeq}{\end{equation}}
\newcommand{\ben}{\begin{eqnarray*}}
\newcommand{\een}{\end{eqnarray*}}

\newcommand{\tauf}{\tau_\mathrm{f}}

\newcommand{\St}{\ensuremath{\mathrm{St}}}

\newcommand{\pderiv}[2]{\frac{\partial{#1}}{\partial{#2}}}
\newcommand{\ptderiv}[1]{\frac{\partial{#1}}{\partial{t}}}
\renewcommand{\v}[1]{{\boldsymbol{#1}}}

\newcommand{\del}{\v{\nabla}}
\newcommand{\grad}{\del}
\newcommand{\Div}{\del\cdot}

\newcommand{\Fig}[1]{Fig.~\ref{#1}}
\newcommand{\fig}[1]{\Fig{#1}}

\newcommand{\Eq}[1]{Eq. (\ref{#1})}

\newcommand{\eq}[1]{\Eq{#1}}

\newcommand{\advec}{\left(\v{u}\cdot\del\right)}
\newcommand{\Table}[1]{Table~\ref{#1}}
\newcommand{\sect}[1]{Sect.~\ref{#1}}

\newcommand{\beqn}{\begin{eqnarray}}
\newcommand{\eeqn}{\end{eqnarray}}
\newcommand{\cv}{c_{_{V}}}
\newcommand{\cp}{c_p}
\newcommand{\taut}{\tau_{_T}}

\graphicspath{{figures/}}
\shorttitle{Planets and photoelectric instability in dust-rich disks}
\shortauthors{Castrejon et al.}
\begin{document}
\title{Disentangling planets from photoelectric instability\\ in gas-rich optically thin dusty disks} 
\author{Areli Castrejon\altaffilmark{1}, Wladimir Lyra\altaffilmark{1,2}, Alexander J.W. Richert\altaffilmark{3}, \& Marc Kuchner\altaffilmark{4}}
\altaffiltext{1}{Department of Physics \& Astronomy, California State University Northridge, 18111 Nordhoff St, Northridge, CA 91330, USA.\\
  areli.castrejon.576@my.csun.edu,wlyra@csun.edu}
\altaffiltext{2}{Jet Propulsion Laboratory, California Institute of Technology, 4800 Oak Grove Drive, Pasadena, CA, 91109, USA.\\
  wlyra@jpl.nasa.gov}
\altaffiltext{3}{Department of Astronomy \& Astrophysics, Penn State University, 525 Davey Lab, University Park, PA 16802, USA.\\
  alexrichert@gmail.com}
\altaffiltext{4}{NASA Goddard Space Flight Center, 8800 Greenbelt Rd, Greenbelt, MD, 20771, USA.\\
  marc.j.kuchner@nasa.gov}

%
%

\begin{abstract}  
  Structures in circumstellar disks such as gaps and rings
  are often attributed to planets. This connection has been
  difficult to show unequivocally, as other processes may also
  produce these features. Particularly, a photoelectric instability (PEI) has
  been proposed, operating in gas-rich optically thin disks, that
  generates structures predicted by planet--disk interactions. We
  examine the question of how to disentangle planetary effects on
  disk structure from the effects of the PEI. We use the Pencil
  Code to perform 2D global hydrodynamical models of the
  dynamics of gas and dust in a thin disk, with and
  without planetary perturbers. Photoelectric heating is modeled
  with an equation of state where pressure is proportional to
  dust surface density. The drag force on grains and its
  backreaction on the gas are included. Analyzing the situation
  without PEI, we find that gas--dust interactions alter the
  shape of the planetary gap from the dust-free case
  when the local dust-to-gas ratio $\varepsilon$ approaches unity.
  This result applies also to primordial disks, because dust
  drifting inwards accumulates at the edge of the planetary
  gap, and any initial dust-to-gas ratio eventually achieves
  $\varepsilon=1$ if the dust reservoir is sufficient. We find a
  result particular to high dust-to-gas ratio disks as well: as
  dust drifts inwards, the dust front becomes a sharp transition,
  and the backreaction triggers the Rossby wave instability. When
  PEI is included, we find that it obscures structures induced
  by planets unless the planet's mass is sufficiently large to
  carve a noticeable gap. Specifically, the instability generates
  arcs and rings of regular spacing: a planet is discernible when
  it carves a dust gap wider than the wavelength of the PEI.
\end{abstract}
%
%
%
%
\section{Introduction}

Exoplanets form and dwell in disks of gas and dust, which show rings,
spirals, and other patterns in optical \citep{Grady+13,Konishi+16,Hornbeck+16}, infrared
\citep{CurrieKenyon09,Benisty+15,Currie+14,Currie+16}, and
sub-millimeter images \citep{Andrews+11,Dong+17a,Muto+12,Dong+15,DongDawson16}.
Planets orbiting in disks perturb them
gravitationally, creating similar patterns to those observed \citep{KleyNelson12} that have in
turn been used to infer the presence of hidden planets 
and constrain their properties 
\citep[e.g.][]{KuchnerHolman03,Chiang+09,Dong+16,DongFung17,DipierroLaibe17,Hord+17}. A planet has been
observed within the gap of the transition disk around PDS~70
\citep{Muller+18,Keppler+18}, a disk that indeed shows several of the
features predicted by planet--disk interaction, validating the approach
of using disk features as signposts for planets. 

Yet, our understanding of disk processes and planet--disk interaction
is not complete. The dynamics of planet--disk interaction are usually modeled
in the dust-free or gas-free limits. The former \citep[e.g.][]{KleyNelson12} is applicable to
primordial disks, where gas dominates the dynamics,
entraining the dust grains seen in optical and infrared images. The
latter \citep[e.g.][]{KuchnerHolman03} is applicable to debris disks, where dust dominates the
dynamics. But as disks evolve from gas-rich toward sparse, gas-free systems like our
own solar system, they inevitably pass through a phase when they
become optically thin, and the gas-to-dust mass ratio drops to the
range of 0.1--1. In this phase,
the disk physics becomes more complex, because dust grains and gas
molecules both interact with stellar photons, and both gas and dust play
an important role dynamically due to their similar total masses. \citep{KlahrLin05,BeslaWu07}. This phase corresponds to young
debris disks and to certain regions of every transitional and pre-transitional
disk.

Over the past two decades, studies have shown that gas indeed is a common component of
many debris disks
\citep[e.g.][]{ChenJura03,Redfield07,France+07,RobergeWeinberger08,Moor+11,WelshMontgomery13,
Riviere-Marichalar+14,Moor+15,Lieman-Sifry+16,Marino+16,Hales+17,Marino+17}.
The bulk of the gas may be metals, or it may be molecular hydrogen, which is much more difficult to detect \citep{Thi+01}. The origin of the gas in these disks is a matter of
continued debate. It is either leftover primordial gas, or it is
produced from solid debris by photodesorption \citep{Grigorieva+07}
or collisions \citep{CzechowskiMann07}, processes that should occur
in every debris disk at some level.

We have only just begun to understand which processes dominate in
gas-rich optically thin disks, the nature of the hydromagnetic
turbulence \citep{KralLatter16}, the instabilities that can arise, 
and how these processes can affect the morphology of 
these disks. As a matter of fact, gap carving by a planet, a problem long
benchmarked \citep{deValBorro+06}, has never been studied 
in the regime of dust-to-gas ratio close to unity. Gap
carving is relatively well understood for gas-rich primordial disks, 
for which the dust back reaction can usually (but not always) be ignored as the
dust-to-gas ratio is close to the interstellar value ($\varepsilon
\approx 10^{-2}$).

In this case, gas is cleared away by planetary
torques, and viscous torques tend to fill the gap \citep[][and
references therein]{KleyNelson12}. It is also
relatively well understood for gas-free debris
disks, as objects are cleared away by eccentricity pumping at
resonances, leaving the planet neighborhood following the so-called
Tisserand tails \citep[e.g.][]{Zhu+14}. Yet, when both gas and dust are dominant, gap carving
may deviate significantly from these idealized cases. This leads to
the first question we pose: what happens to a disk gap when dust is added to the point that the gas drag
backreaction changes the torques acting to close the gap?

A second question we pose, that depends on understanding the first, is
related to the combined effect of the gravitational potential of a
planet and a dynamical instability in gas-rich optically
thin disks. The instability, suggested by
\cite{KlahrLin05,BeslaWu07}, is affected by the
combined operation of two dynamical
processes. First, if the disk is optically thin, stellar UV photons eject 
electrons from dust grains, heating the gas
\citep{KampvanZadelhoff04,Jonkheid+04,Zagorovsky+10}. Second, the
hot gas provides a local pressure maximum that concentrates the dust grains 
\citep{Whipple72}, establishing a positive feedback process. In just a few orbital periods,
this {\it photoelectric instability} (PEI) can radically reshape the
disk structure. In the first 2D and 3D models of this instability,
\cite{LyraKuchner13} found that
the PEI would concentrate power in high wavenumbers if not for the
effect of the gas drag backreaction, which organizes the dust in rings
and arcs. They also found that the PEI leads to several patterns usually associated with planet--disk
interactions, challenging any correlations drawn between dust
structures and perturbing planets. \cite{Richert+18} incorporate radiation pressure into a PEI model,
and show that while grains of different sizes feel very different
amounts of different radiation pressure 
force, the PEI can still produce dramatic structures in such a more realistic disk model,  
depending on the parameters of the system. The morphology of the flow depends strongly on the gas
density (which determines the gas drag), the dust density (which
determines the photoelectric heating), and the dust-to-gas ratio (which
determines the drag backreaction). For disks with
modest level of gas $\approx 10^{-6} \ {\rm g\,cm^{-2}}$ and dust-to-gas
ratio $\varepsilon=10^{-1}$, sharp concentric rings and arcs are seen, showing
the anticorrelation between gas and dust seen in models of the PEI without
radiation pressure \citep{LyraKuchner13}. For higher surface density $\approx 10^{-4} \ {\rm
  g\,cm^{-2}}$ and lower gas-to-dust ratio $\varepsilon=10^{-3}$, erratic spiral structure is seen
throughout the disk, with transient small-scale structure in dynamical
timescales, and no obvious tendency of the dust and gas to
anticorrelate. At even higher gas densities, outward gas flows can
entrain even larger grains that do not participate in the
instability.  

Given the variety of structures that the PEI can generate in a
  disk, which muddles an unequivocal association with planetary
  perturbations, a natural question to ask is: what are the expected patterns
  when both a planet and PEI are present in a disk? How do we
  disentangle them? This is the goal of the current study.
 
We structure this paper as follows. In \sect{sect:method} we describe
the equations and numerical method used. In \sect{sect:results} we
present our results, concluding in \sect{sect:conclusions}. 

\section{Method}
\label{sect:method}
%
%
\subsection{The Equations}
We work in the thin disk approximation using the vertically averaged equations of hydrodynamics. Our model includes gas and dust; the gas is evolved in an Eulerian grid, whereas the dust is treated as Lagrangian particles. The equations solved are
\beqn
\pderiv{\varSigma_g}{t} &=& -\advec{\varSigma_g} -\varSigma_g \Div{\v{u}}, \label{eq:continuity} \\ 
\pderiv{\v{u}}{t} &=& -\advec{\v{u}} -\frac{1}{\varSigma_g}\grad{P} - \grad{\varPhi} - \frac{\varSigma_d}{\varSigma_g}\frac{\left(\v{u}-\v{v}\right)}{\tauf}, \label{eq:euler}\\
\pderiv{S}{t} &=& -\advec{S} -\frac{c_V}{T}\frac{\left(T-T_p\right)}{\taut}, \label{eq:entropy}\\
\varPhi &=& \sum_i^n\frac{GM_i}{\sqrt{{R_i}^2 + {b_i}^2}}, \label{eq:potential}\\
P &=& \rho {c_s}^2/\gamma, \label{eq:eos}\\
\frac{d\v{x}}{dt} &=& \v{v}, \label{eq:position-dust}\\
\frac{d\v{v}}{dt} &=& -\grad{\varPhi} - \frac{\left(\v{v}-\v{u}\right)}{\tauf} \label{eq:momentum-dust},
\eeqn
In the equations above, $\varSigma_g$ and $\varSigma_d$ are the vertically
integrated gas density and bulk density of dust, respectively. Their 
quotient defines the dust-to-gas ratio $\varepsilon=\varSigma_d/\varSigma_g$.
{\footnote{As we use the 2D infinitely thin approximation, the different scale heights 
of grains and gas is not resolved. The definition of dust-to-gas ratio used here is the 
ratio of their {\it column} densities; in 3D it should be the ratio of their 
{\it volume} densities, which, because of the different degree of stratification, 
is larger for loosely coupled grains in the midplane, by $\exp{[(\St+\delta)/\delta}]$ 
\citep{LyraLin13}. For well-sedimented grains, the 2D approach may underestimate the local 
dust-to-gas ratio by orders of magnitude.}}
The velocity of gas parcels is given by 
$\v{u}$ ; $P$ stands for the vertically integrated pressure,
$\varPhi$ is the gravitational potential, and $\v{v}$ is the velocities of dust grains.
The gravitational potential includes contributions
from the star and embedded planets, treated as massive particles with a N-body
evolution dynamically integrated with the hydrodynamics evolution.  In \eq{eq:potential}
$G$ is the gravitational constant, $M_i$ is the mass of particle $i$ and $R_i = |\v{r} - \v{r}_{p_i}|$
is the distance relative to particle $i$, located at $\v{r}_{p_i}$. The quantity $b_i$
is the distance over which the gravity field of particle $i$ is softened to prevent singularities.
In \eq{eq:entropy}, $S$ is the gas entropy, given by
\beq
S = \cv\left[ \ln\left(\frac{P}{P_0}\right) - \gamma \ln\left(\frac{\varSigma_g}{\varSigma_{g0}}\right)\right]
\eeq
\noindent where $\cv$ is the specific heat at constant volume, $\gamma=\cp/\cv$ is the
adiabatic index, and $c_P$ is the specific heat at constant pressure; $P_0$ and
$\varSigma_{g0}$ are reference pressure and gas density values, respectively.
In \eq{eq:entropy} $T$ stands for the gas temperature
\beq
T = \frac{c_s^2}{\cp\left(\gamma-1\right)}
\eeq
\noindent where $c_s$ is the sound speed; $\taut$ is the
thermal coupling times between gas and dust. The quantity $T_p$ is  the gas 
temperature set by equilibrium between the different heating and cooling 
processes, including photoelectric heating. Calculating it accurately 
necessitates detailed radiative transfer that is beyond the scope 
of the present work. Instead, we use a simple
prescription for the gas temperature set by photoelectric heating as used in \cite{KlahrLin05}
\beq
T_p = T_0\left(\frac{\varSigma_d}{\varSigma_{g0}}\right)^\beta \label{eq:reftemperature}\\
\eeq
\noindent where $T_0$ is a reference temperature. We keep $\beta=1$ 
 (see the appendix of \citealt{KlahrLin05} and Fig 1 of that work 
for the estimate of $\beta\approx 1$ given by a balance between photoelectric heating 
off silicate grains and cooling via the C{\sc II} line.) In the Lagrangian
dust equations, $\v{v}_p$ is the velocity of a dust grain, and $\v{x}_p$ is their
position. The last term in \eq{eq:momentum-dust} is the drag force by which
gas and dust interact dynamically. It brings the velocities of the dust grains
to that of the gas within the timescale $\tauf$,  the friction time. The last term
in \eq{eq:euler} is the backreaction of the drag force. We choose the friction time
as proportional to the local orbital period
\beq
\tauf = \tauf{_0}\left(\frac{\varOmega_0}{\varOmega}\right)
\label{eq:tauf-omega}
\eeq
\noindent  as in this case the relative timescales (the PEI growth rate also 
scales with the orbital period) are similar at all radii{\footnote{The drag time in the Epstein regime is \citep[cf.][]{Youdin10} $\tau_f = a_\bullet \rho_\bullet /(c_s \rho)$, where  $a_\bullet$ is the grain radius, $\rho_m$ is the grain's material density, and $\rho$ the gas volume density. As such, \eq{eq:tauf-omega} is equivalent to grains of identical size and material density scattered through a disk where $c_s\rho \propto \varOmega$. This would be the case for, e.g., $\rho \propto 1/r$ and $c_s \propto 1/\sqrt{r}$, which are not unreasonable choices for the density and temperature gradients.}}. We keep $\tauf{_0}\varOmega_0 = 1$.  The operator
\beq
\frac{D}{Dt} = \ptderiv{} + \v{u} \cdot \del
\eeq
\noindent represents the advective derivative.

Given that the goal of this work is to examine the competing roles of the PEI and gravitational perturbation by a planet, we ignore the effects of radiation pressure. Because the thermal time is expected to be very low, we use the instantaneous thermal coupling approximation \citep{LyraKuchner13}. That means that instead of solving the energy equation, and equate $T = T_p$ and update the sound speed is updated accordingly. This change effectively amounts to choosing a new equation of state that depends on the dust density
\beq
\lim_{\taut\to 0} P = \cv T_0 (\gamma-1) \ \left(\frac{\varSigma_g}{\varSigma_{g0}}\right) \varSigma_d
\eeq

We solve the equations with the {\sc Pencil Code}{\footnote{The code, including improvements done for the present work, is publicly available under a GNU open source license and can be downloaded at http://www.nordita.org/software/pencil-code}}, which integrates the evolution equations with sixth order spatial derivatives, and a third order Runge--Kutta time integrator. Sixth-order hyperdissipation terms are added to equations of motion, to provide extra dissipation near the grid scale. They are needed because the high order scheme of the Pencil Code has little overall numerical dissipation \citep{McNally+12}. The calculations are sped up by using the ``fargo'' orbital advection algorithm \citep{Masset00,Lyra+17}.
%
%

\begin{figure}
  \begin{center}
    \resizebox{\columnwidth}{!}{\includegraphics{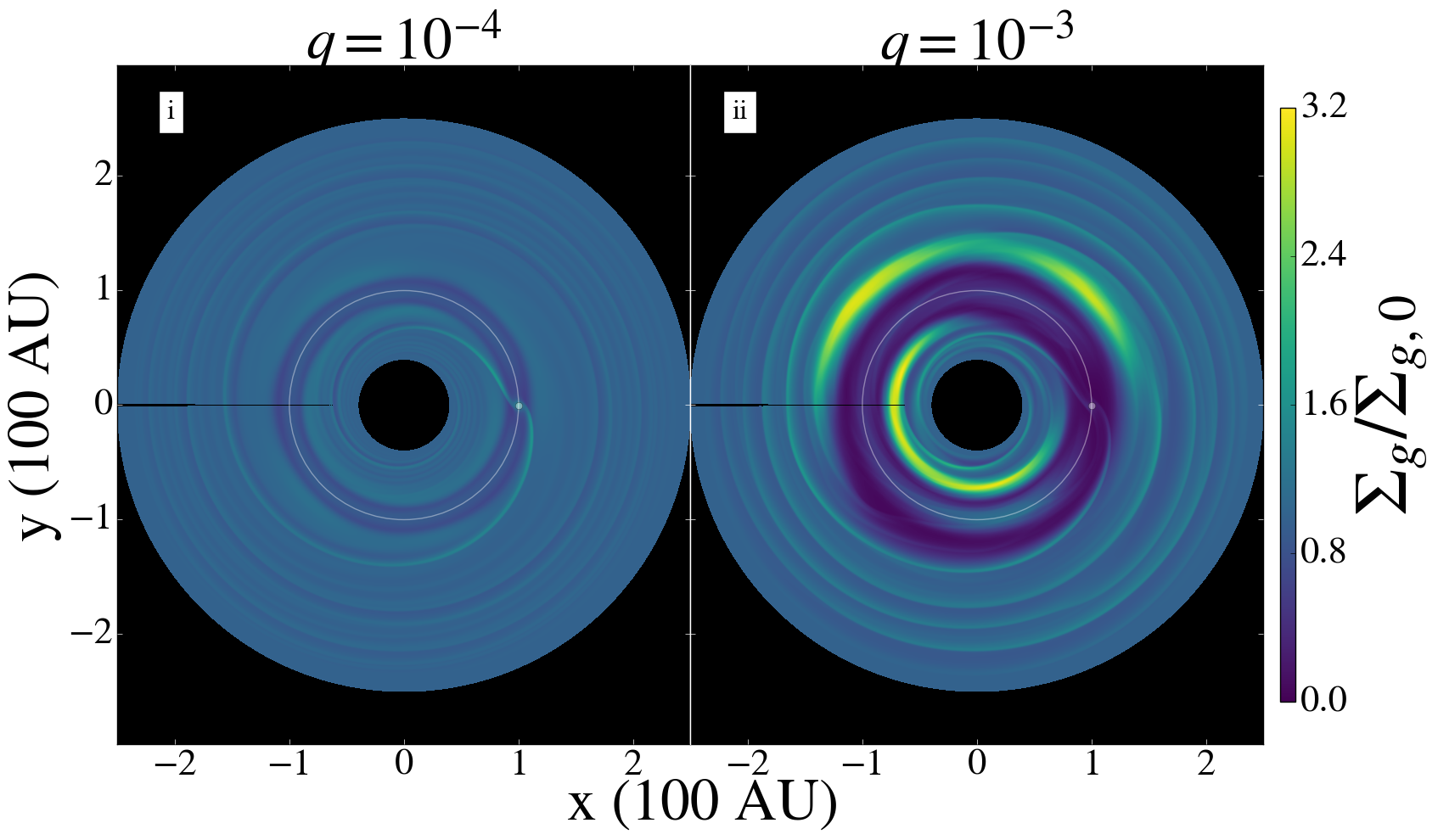}}
    \resizebox{\columnwidth}{!}{\includegraphics{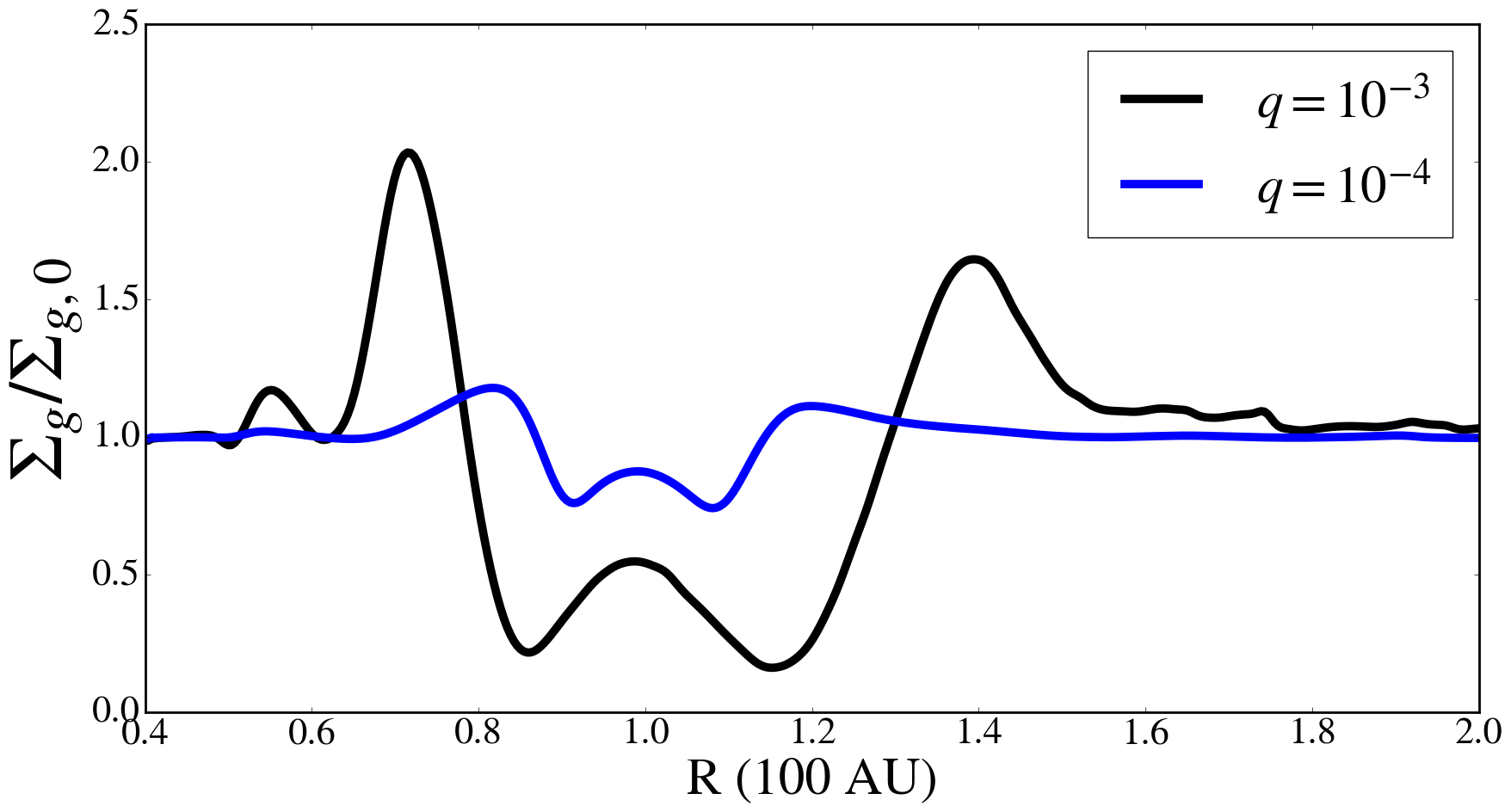}}
    \caption{Fiducial run of a global disk containing only
      gas. \textit{Upper left:} Planet of mass ratio
      $q=10^{-4}$. \textit{Upper right:} Planet of mass ratio
      $q=10^{-3}$. Both planets carve a gap and generate a spiral in
      the gas. \textit{Bottom:} Averaged gas density as a function of
      radius from the star. The black curve represents a planet with
      mass ratio $q=10^{-3}$ and the blue curve represents
      $q=10^{-4}$. These models represent benchmarks against which we
      will compare models including dust and photoelectric instability.}
    \label{fig:devalborro}
 \end{center}
\end{figure}

\begin{figure*}
  \begin{center}
    \resizebox{.9\textwidth}{!}{\includegraphics{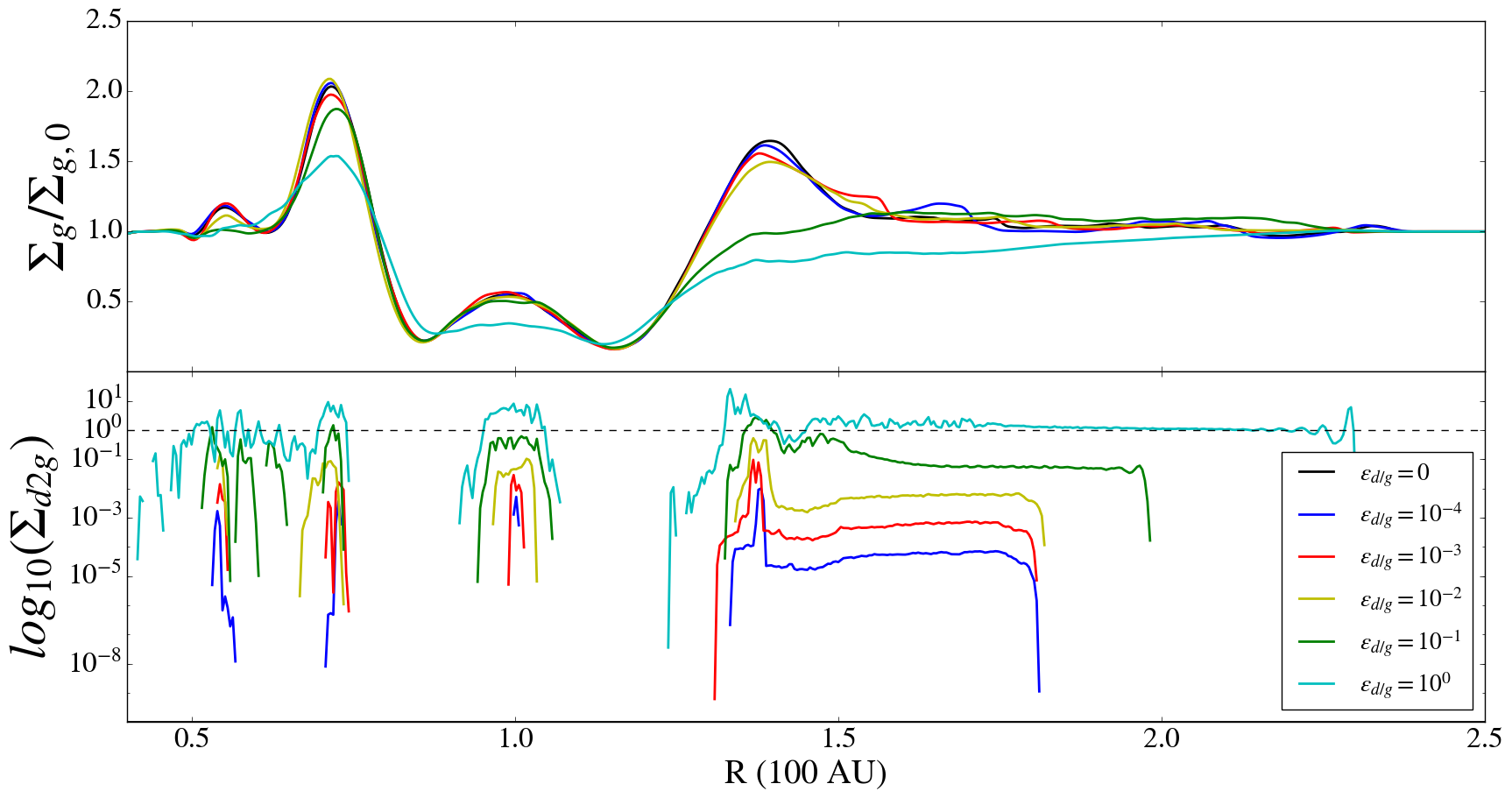}}
  \end{center}
  \caption[]{Effect of dust-to-gas ratio on the shape of planetary gaps. Snapshots are
    taken at 100 disk orbits for all cases. The top panel shows the
    gas gap shapes for different initial dust-to-gas ratios, for a
    Jupiter mass planet. The bottom panel shows the local dust-to-gas
    ratios. In the upper panel the black line shows a dustless
    model. We see that as the dust-to-gas ratio increases, 
    the pressure bump in the inner gap edge eventually gets significantly
    depressed, and the bump in the outer gap edge is almost largely
    washed out. Inspecting the lower plots, we notice that these significant deviations from the dustless profile start to
    become prominent when the local dust-to-gas ratio approaches
    unity. This is expected, as the backreaction of the dragforce
    comes to heavily influence the gas motion. This threshold is
    easily reached in the pressure bumps of the planetary gap edges
    for the runs with initially high dust-to-gas ratios, but we stress
    that even for $\varepsilon=10^{-2}$ (and lower) the accumulation in the gap 
    edge will eventually reach unity, if the drift of pebbles is
    continuous and the disk is large enough. We also remark that
    the disk initially did not have a global negative pressure
    gradient. The resulting particle drift is effected by the
    planetary gap. A global negative pressure gradient will speed up the
    process.  Finally, we highlight that because this simulation suite
    does not include photoelectric heating, this particular
    result is not restricted to optically thin disks, and thus applies
    also to primordial and transition disks.}
  \label{fig:qe-3gasd2g}
\end{figure*}

\begin{figure*}
  \begin{center}
    \resizebox{.9\textwidth}{!}{\includegraphics{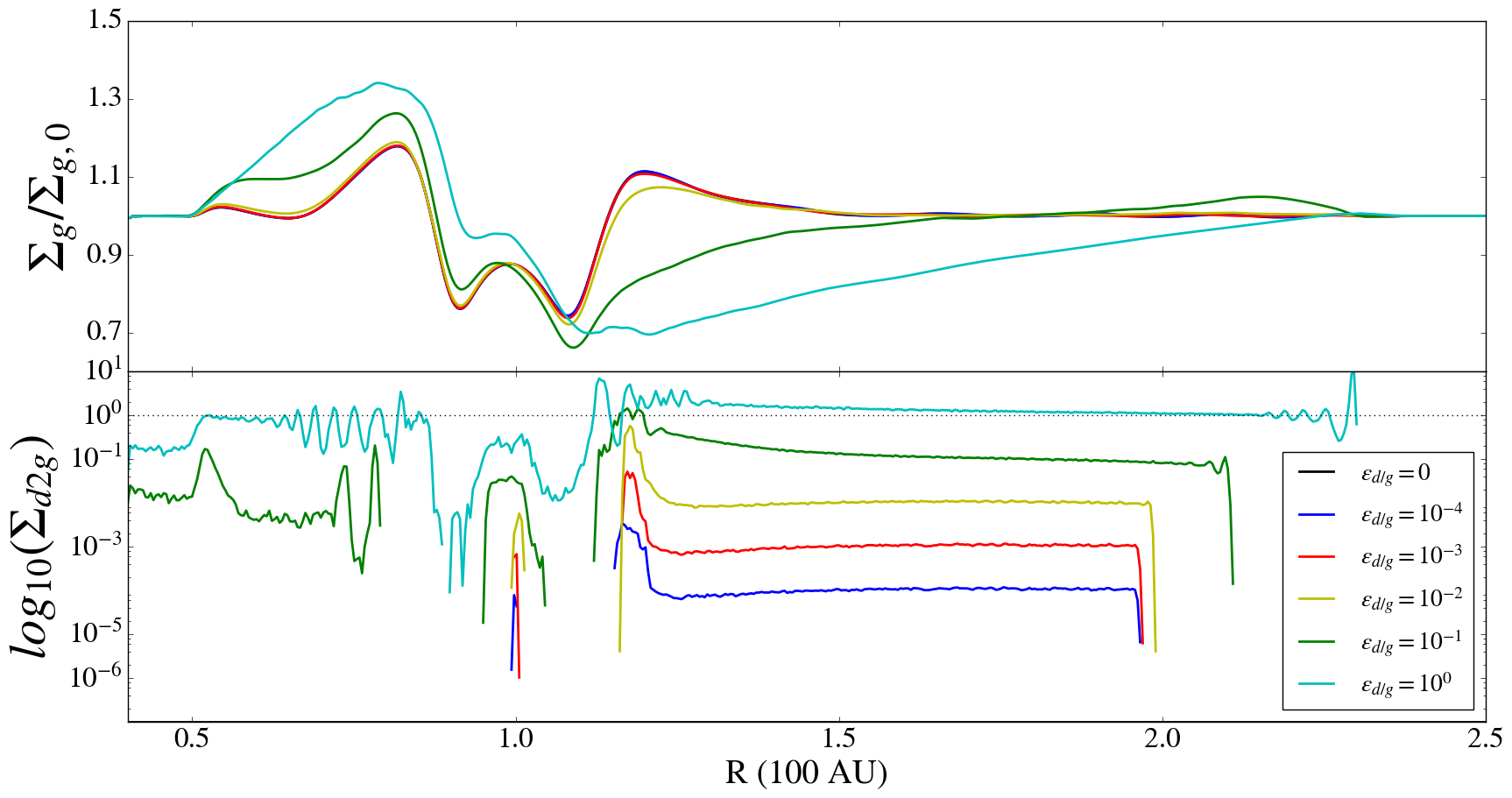}}
  \end{center}
  \caption[]{Same as \fig{fig:qe-3gasd2g}, but for $q=10^{-4}$.  A high dust
    to gas ratio value once again causes the gap shapes to be
    distorted, strongly for the cases of initial $\varepsilon=1$ and
    $0.01$, but also for the $\varepsilon=0.01$ case.  Curiously, the effect in 
    the inner bump is the opposite of the higher mass planet
    $q=10^{-3}$. We explain it in \sect{sect:innerdisk} and \fig{fig:drift}.}
  \label{fig:figneptune}
\end{figure*}

\begin{figure}
  \begin{center}
    \resizebox{\columnwidth}{!}{\includegraphics{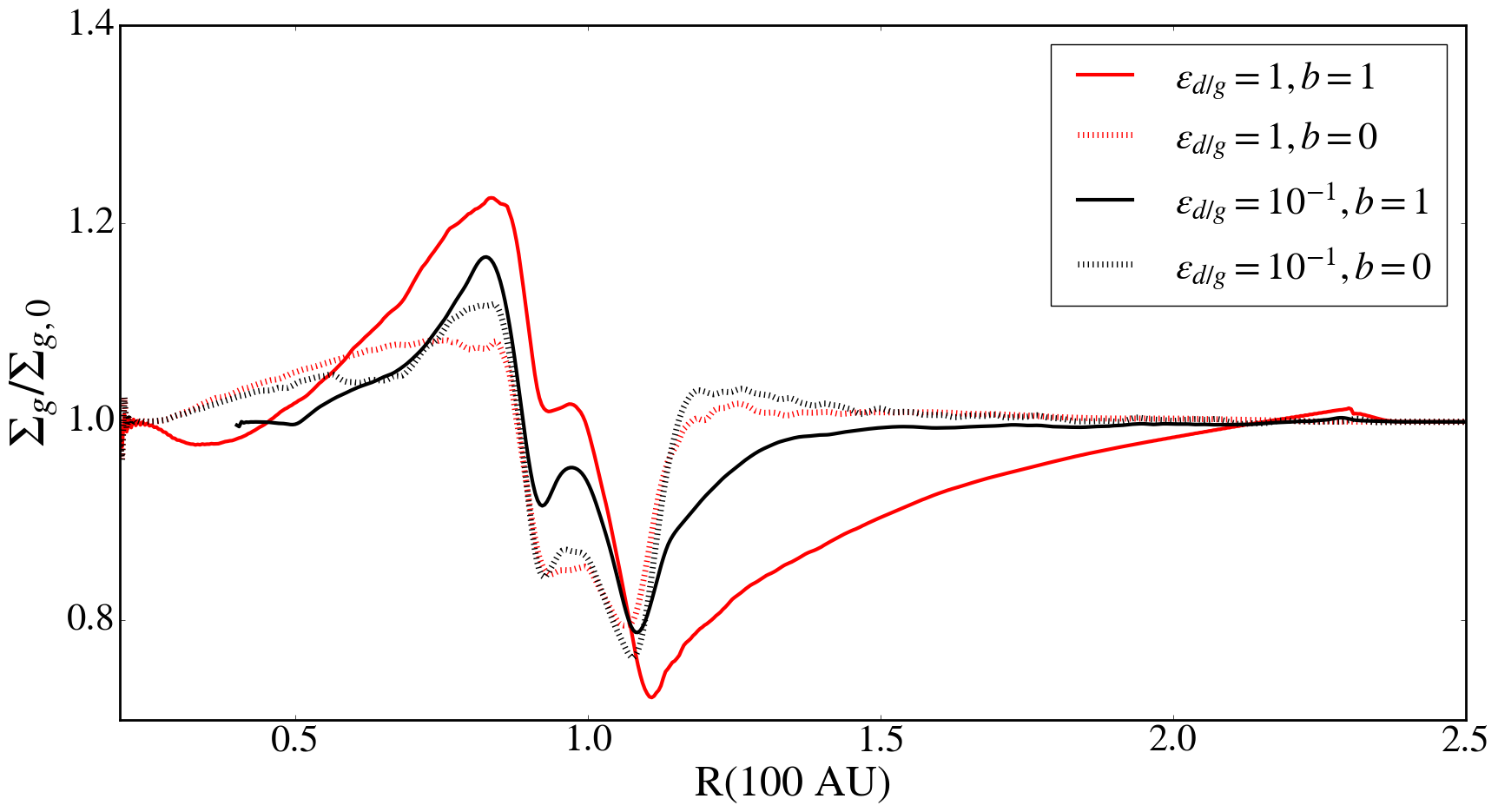}}
  \end{center}
  \caption[]{The effect of the particle drift and angular
    momentum conservation in the gap profile of a
    Neptune-mass planet. The red lines show runs with
    $\varepsilon=1$. The black lines show runs with $\varepsilon=0.1$. Solid lines have $b=1$ ($b$ is
    the power law of the temperature, giving a global negative pressure
    gradient); the dashed lines are control runs correspond to $b=0$ and thus no global
    negative pressure. The control runs are similar, with the
    higher-dust to gas ratio showing a small depression. However, when
    drift is included, to conserve angular momentum the gas is pushed
    outwards. For higher dust-to-gas ratio (solid red line),
    more gas is pushed outwards than for lower dust-to-gas ratio (black
    solid line).}
  \label{fig:drift}
\end{figure}

\subsection{Run Parameters and Units}
The simulations are performed on a uniform cylindrical grid (coordinates $r,\phi$) with
resolution $(N_r,N_{\phi}) = (528,528)$ and linear domain $L_r$=[0.4,2.5] and full 2$\pi$
coverage in azimuth. To represent the dust grains, 200,000 Lagrangian superparticles are scattered uniformly
across the disk. The star and the planet orbit around the barycenter of the system in
circular orbits. Disk self-gravity is ignored. We ran simulations with three different planet masses. The mass ratio of the
planet to the host star are $q=10^{-3}, 10^{-4}, 10^{-5}$. For each
planet we simulated 5 different dust-to-gas ratios, from $10^{-4}$ to 1,
logarithmically spaced. The runs from  $\varepsilon=10^{-2}$ to
$\varepsilon=1$ are in the zone of highest growth rate for the PEI
\citep{LyraKuchner13, Richert+18}. 
We choose our units such that
\beq
GM_\star = \varOmega_0 = r_0 = 1
\eeq
All quantities with subscript ``0'' are measured at $r_0$. For $r_0=100\,AU$ and
$M_\star=1M_\odot$, our unit of time is $\approx$ 160 yr, and the unity of velocity
$\approx$ 3\,km\,s$^{-1}$. We investigate disks of aspect ratios $h \in \{0.05, 0.1, 0.2\}$, where $h=H/r$, with $H=c_s/\varOmega$ the disk scale height. For a solar mixture of
molecular hydrogen and helium, of mean molecular weight $\mu=2.5$ and $\gamma=7/5$, these correspond
to temperatures of $T$ = 5, 20, and 80\,K, respectively.
We consider monodispersive particles of radius $a_\bullet=1\mu$m. This choice, along with the choice of Stokes number unity
\beq
\St \equiv \tauf \varOmega = 1
\eeq
\noindent sets the unit of density {\footnote{The Stokes number
      should in fact be dynamic, updated by density and temperature as
      given by \eq{eq:tauf}. Setting it as a constant is a choice we
      make, to better isolate the physical phenomenon we want to
      study. Yet, notice that because the photoelectric heating introduces an
      anti-correlation between gas density and gas temperature, the product
      $c_s\rho$ varies a lot less than either $c_s$ or $\rho$, so $\St
      \approx {\rm const}$ is not
      a bad first approximation).}}. This is because the friction time in the Epstein regime is
\beq
\tauf = \frac{a_\bullet \rho_\bullet}{\rho c_s},
\label{eq:tauf}
\eeq
\noindent where $\rho_\bullet$ is the grain internal density. For icy grains, it
should be $\rho_\bullet$ $\approx 1\,$g\,cm$^{-3}$. Multiplying \eq{eq:tauf} by
$\varOmega$ to obtain the Stokes number and setting to one, yields, solving for density
\beq
\rho \, \vert_{\St=1} = \frac{a_\bullet \rho_\bullet}{c_s}
\eeq
\noindent which results, for $h=0.1$, in $\rho \approx 10^{-8}$ g\,cm$^{-3}$, with associated column density of $\varSigma=\sqrt{2\pi} \rho H \approx 2.5\times 10^{-4}$ g\,cm$^{-2}$.
We keep the gas and dust densities constant with radius, whereas
the temperature decreases according to $T=T_0 r^{-b}$. We usually
use $b=1$ corresponding to a negative global pressure gradient,
causing the dust to slowly drift inwards. Although a drift is
present, the resolution is not high enough to capture the streaming
instability \citep{YoudinGoodman05,Johansen+07,LyraKuchner13}. A subset of simulations
use $b=0$.

The parameters of the simulations we ran are shown in
Table~\ref{table:allmodels}. We explored the parameter space of planet mass, sound speed, and dust-to-gas ratio.
\begin{table}
  \caption[]{Parameter space explored in our simulations: $h$ is the
    disk scale height of the disk, $\varepsilon$ is the initial
    dust-to-gas ratio, and $q$ is the planet-to-star mass ratio.
    The letter identifier in the first column is shown in the upper 
      left corner of each plot in this paper, for easier referral.}
  \label{table:allmodels}
  \begin{center}
    \begin{tabular}{lcc c} \hline \hline
      \multirow{2}{*} \sc{Run} & h  & $\varepsilon$ & $q$ \\
      \multicolumn{4}{c}{\multirow{2}{*}{Fiducial runs with no dust}}\\\\\hline
      i  & 0.05 & $0$ & $10^{-3}$\\
      ii & 0.05 & $0$ & $10^{-4}$\\\hline
      \multicolumn{4}{c}{\multirow{2}{*}{Runs with a single planet.}}\\\\\hline
      A & 0.05 & $10^{-1}$ & $10^{-3}$\\
      B & 0.05 & $10^{-1}$ & $10^{-4}$\\
      F & 0.05 & $1$       & $10^{-3}$\\
      G & 0.05 & $1$       & $10^{-4}$\\
      K & 0.1  & $1$       & $10^{-3}$\\
      L & 0.1  & $1$       & $10^{-4}$\\
      P & 0.2  & $1$       & $10^{-3}$\\
      Q & 0.2  & $1$       & $10^{-4}$\\
      AA & 0.05 & $10^{-4}$ & $10^{-3}$\\
      AB & 0.05 & $10^{-4}$ & $10^{-4}$\\
      AC & 0.05 & $10^{-3}$ & $10^{-3}$\\
      AD & 0.05 & $10^{-3}$ & $10^{-4}$\\
      AE & 0.05 & $10^{-2}$ & $10^{-3}$\\
      AF & 0.05 & $10^{-2}$ & $10^{-4}$\\\hline
      \multicolumn{4}{c}{\multirow{2}{*}{Runs with photoelectric instability.}}\\\\\hline
      C & 0.05 & $10^{-1}$ & 0\\
      H & 0.05 & $1$       & 0\\
      M & 0.1  & $1$       & 0\\
      R & 0.2  & $1$       & 0\\\hline
      \multicolumn{4}{c}{\multirow{2}{*}{Runs with a planet and instability.}}\\\\\hline
      D & 0.05 & $10^{-1}$ & $10^{-3}$\\
      E & 0.05 & $10^{-1}$ & $10^{-4}$\\
      I & 0.05 & $1$       & $10^{-3}$\\
      J & 0.05 & $1$       & $10^{-4}$\\
      N & 0.1  & $1$       & $10^{-3}$\\
      O & 0.1  & $1$       & $10^{-4}$\\
      S & 0.2  & $1$       & $10^{-3}$\\
      T & 0.2  & $1$       & $10^{-4}$\\\hline
      \multicolumn{4}{c}{\multirow{2}{*}{Runs with no gas backreaction.}}\\\\\hline
      BA & 0.05 & $10^{-1}$ & 0\\
      BB & 0.05 & $10^{-2}$ & 0\\\hline
    \end{tabular}
  \end{center}
\end{table}

To investigate the effect of the photoelectric instability we ran 36
simulations. We used two planets, analogous to a Neptune and a Jupiter
mass planet. For each planet we varied 3 disk temperatures, 2
dust-to-gas ratios, and 3 different cases for carving a gap in the
disk. The cases are: the effect of a single planet with no
photoelectric instability, a disk with the instability on its own, and
a combination of an embedded planet and the instability. 

\begin{figure*}
  \begin{center}
    \resizebox{.8\textwidth}{!}{\includegraphics{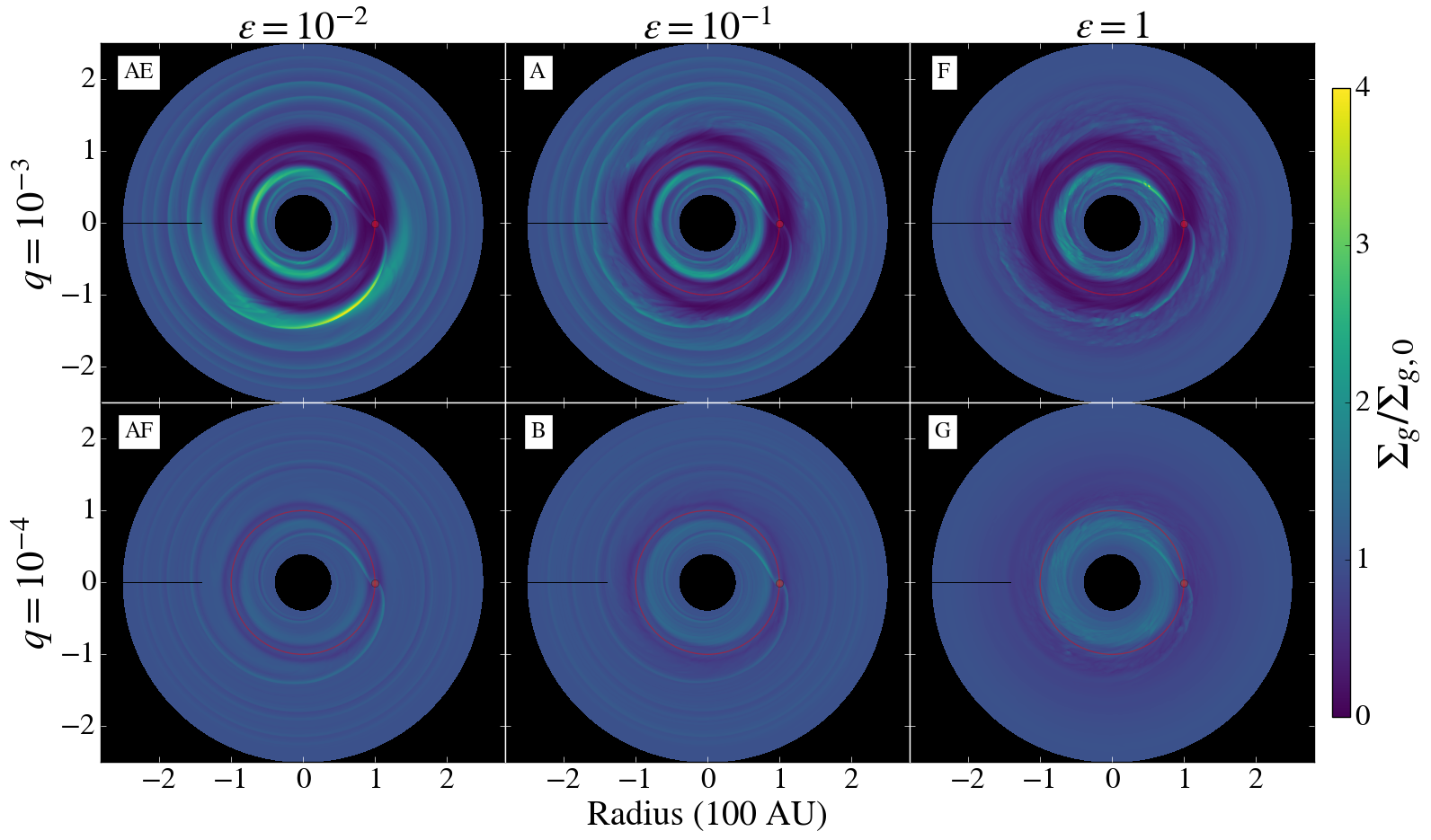}}
  \end{center}
  \caption{Two dimensional visualization of the effect of dust-to-gas ratio shown
    in \fig{fig:qe-3gasd2g} and \fig{fig:figneptune}. The plots show the gas
    density for a disk with a planet of mass ratio $q=10^{-3}$
    (upper plots) and a planet of mass ratio $q=10^{-4}$ (lower plots), with dust-to-gas ratio $\varepsilon=10^{-2}$ (left plots),
    $10^{-1}$ (middle plots), and 1 (right plots). In the case of $\varepsilon=10^{-2}$ the
    dust concentration does not get high enough to significantly
    affect the gas dynamics, and the situation looks very similar 
    to the $\varepsilon=0$ case, shown in \fig{fig:devalborro}. However, 
    for $\varepsilon=0.1$ the effect is visible on gap edges, on 
    the gas in corotation, and on the spiral wake, that all acquire a
    fuzzy and turbulent-like appearance. In the $\varepsilon=1$ 
    case the spiral wake break ups into eddies akin to cigarette smoke.}
  \label{fig:epsilongap}
\end{figure*}
%
%
\section{Results}
\label{sect:results}
\begin{figure*}
  \begin{center}
    \resizebox{.8\textwidth}{!}{\includegraphics{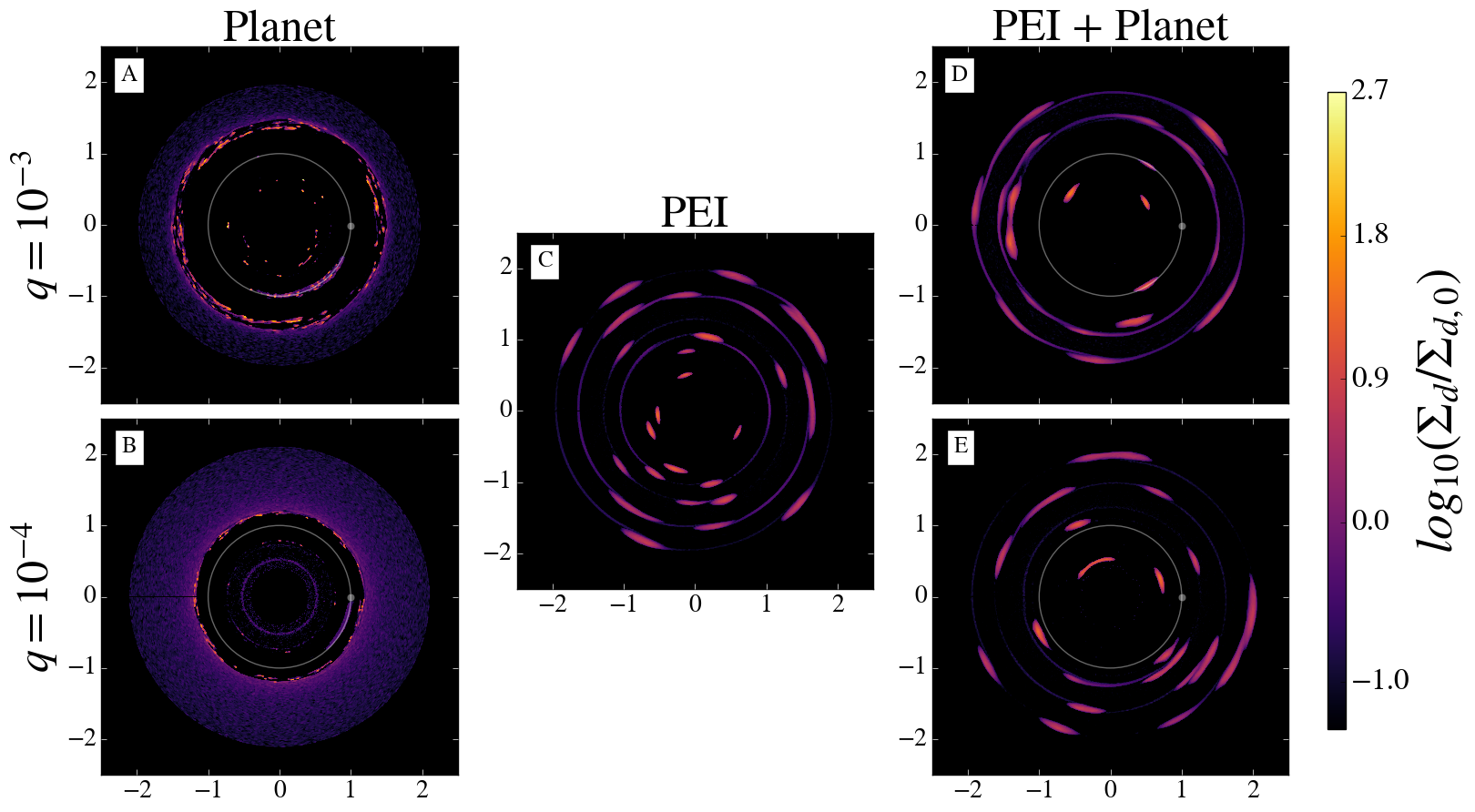}}
    \resizebox{.8\textwidth}{!}{\includegraphics{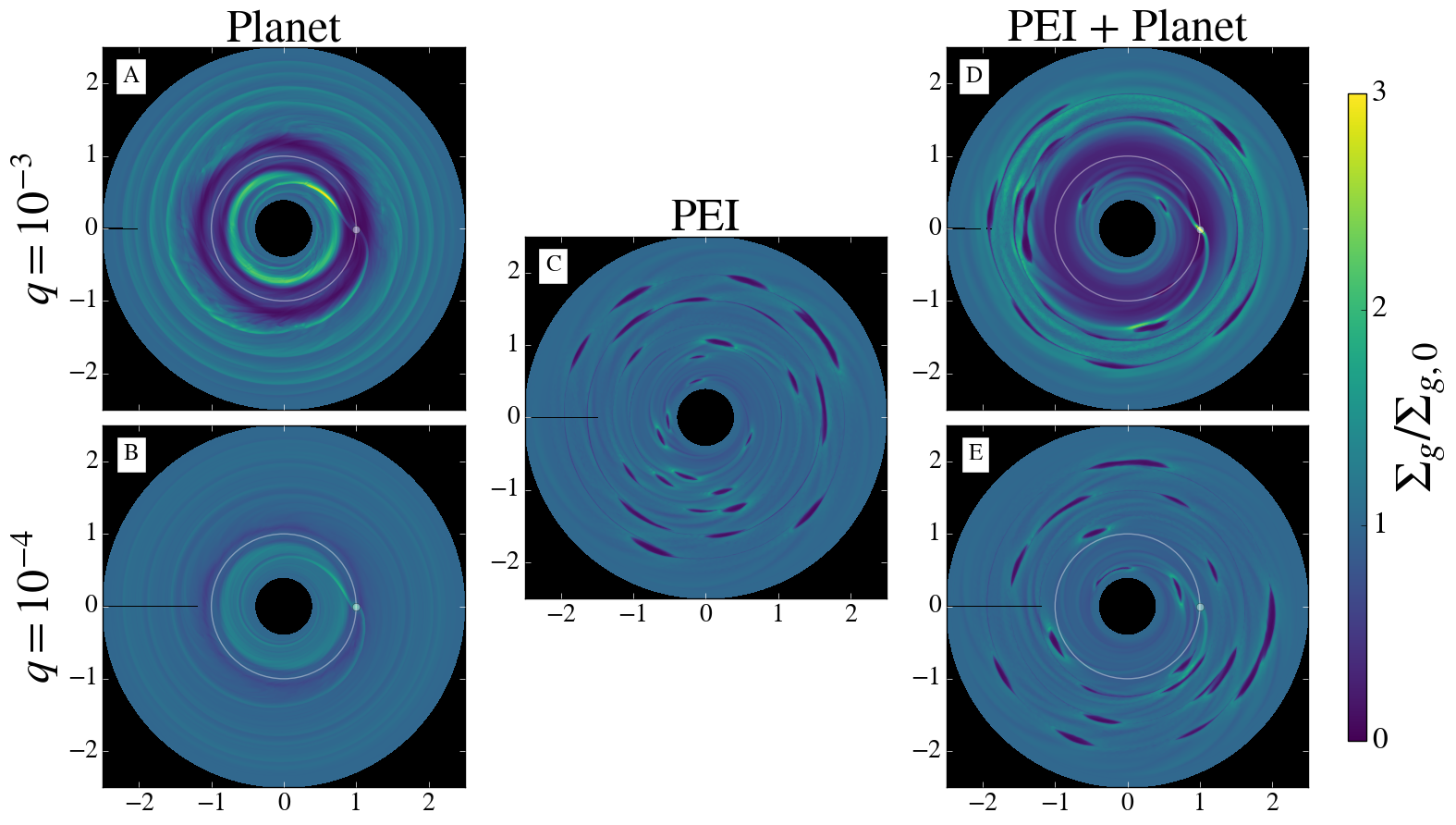}}
  \end{center}
  \caption{Dust (upper panels) and gas (lower panels) in
    a disk with initial dust-to-gas ratio $\varepsilon=10^{-1}$. The
    top row shows results for a disk with a planet of mass ratio $q=10^{-3}$ and the bottom
    row shows results for mass ratio $q=10^{-4}$. The first column
    shows the effect of planet with no photoelectric heating. The
    middle column, labeled ``PEI'', shows the effect of the
    photoelectric instability on its own. The last column shows the
    combined effect of photoelectric instability and a planet. The
    most striking feature of this comparison is that the $q=10^{-4}$ 
    planet with PEI is virtually indistinguishable from the PEI alone
    (the middle case). The main feature that makes the $q=10^{-3}$
    distinguishable in the midst of PEI structures is the deep and
    clear planetary gap. We conclude that when photoelectric heating is included, we find that
      the photoelectric instability obscures structures induced by planets
      unless the planet's mass is sufficiently large to carve
      a noticeable gap. The plots are shown after 100 planetary orbits.}
  \label{fig:eps01dustgas}
\end{figure*}

\begin{figure*}
  \begin{center}
    \resizebox{.45\textwidth}{!}{\includegraphics{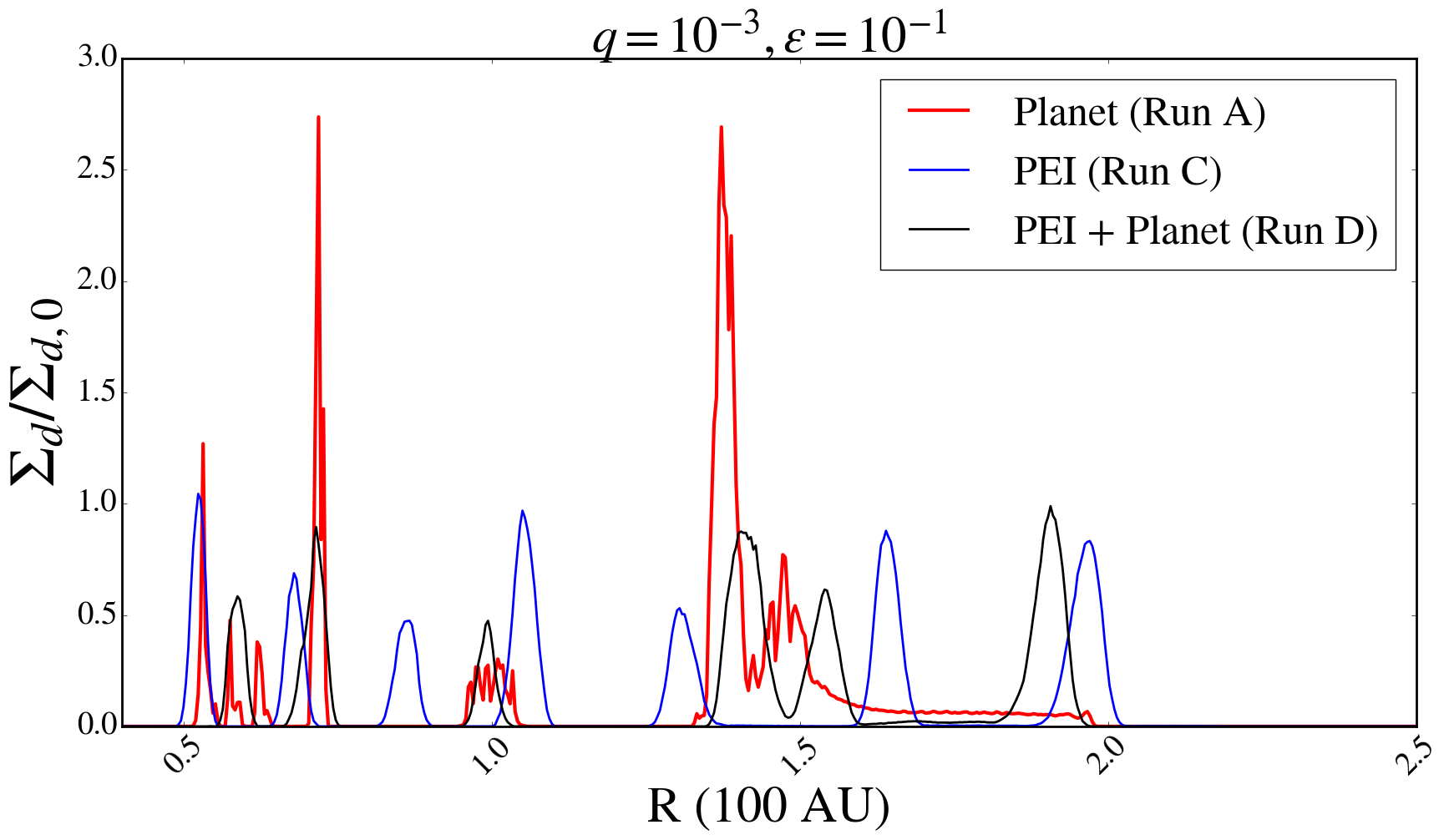}}
    \resizebox{.45\textwidth}{!}{\includegraphics{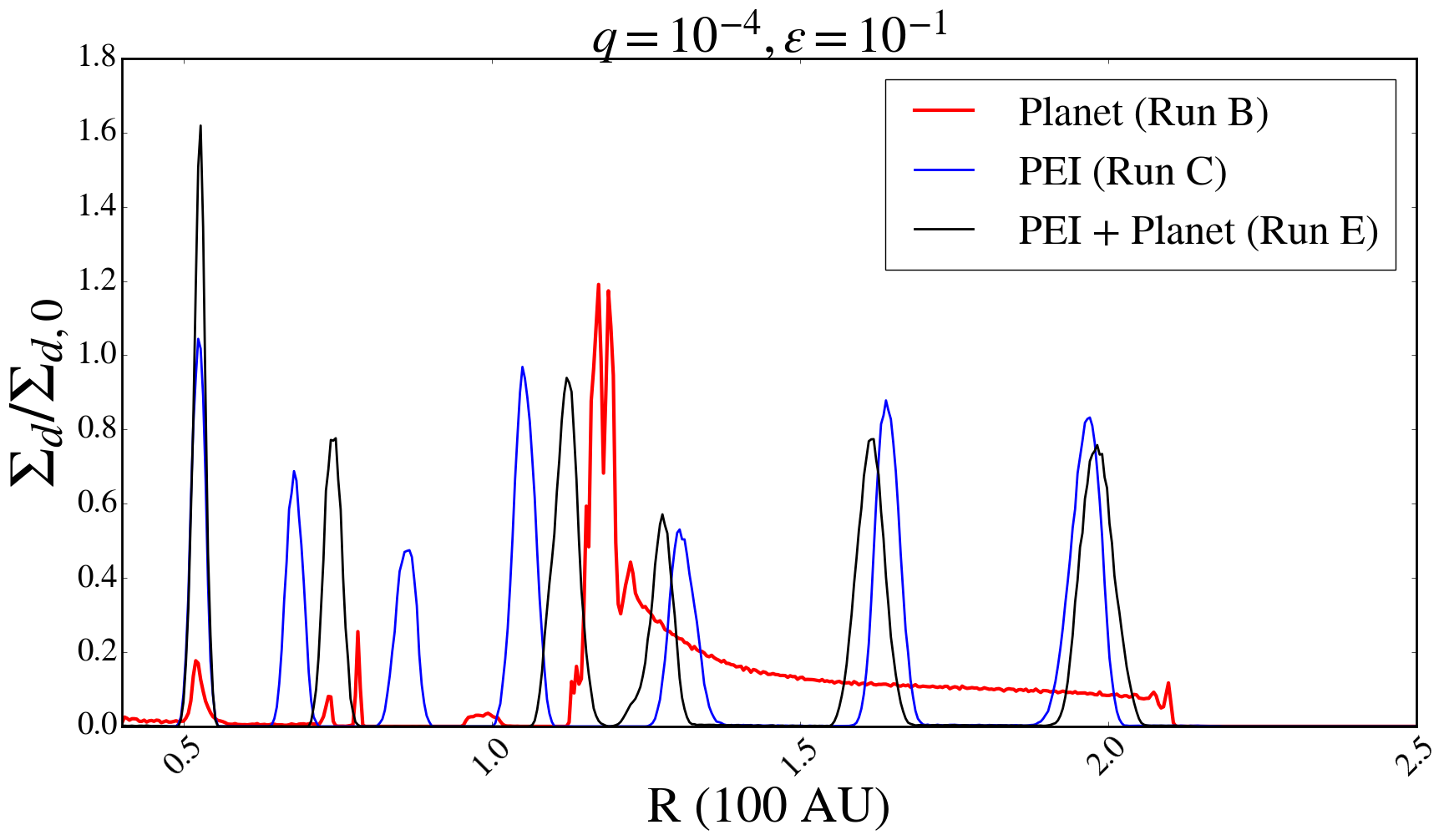}}
  \end{center}
  \caption{Azimuthal average of the dust density for the simulations shown in
    \fig{fig:eps01dustgas}. The left plot shows the $q=10^{-3}$ case, the
    right plot shows the $q=10^{-4}$ case. The cases of planet only
    (red line) and planet + instability (black line) both have a large gap in the dust, centered at $r=1$, where the
    planet is located. The pure PEI case (blue line) shows structure inside the
    predicted location of the gap. The lower mass planet also carves a
    dust gap, but the pure instability and combined case coincide
    almost exactly. The planet cannot easily be disentangled 
    from the instability in this case.}
  \label{fig:dustprofile}
\end{figure*}

\begin{figure*}
  \begin{center}
    \resizebox{.8\textwidth}{!}{\includegraphics{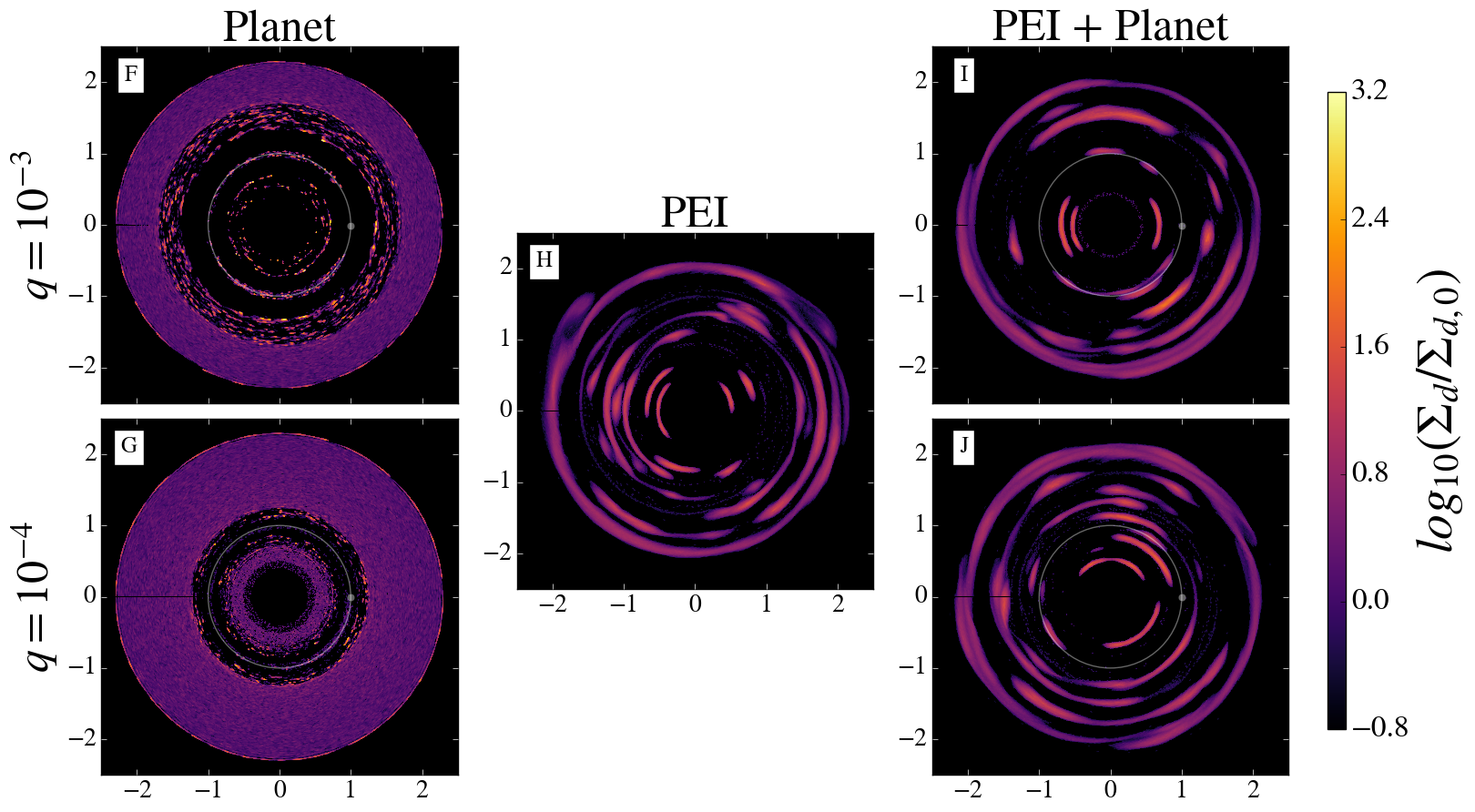}}
    \resizebox{.8\textwidth}{!}{\includegraphics{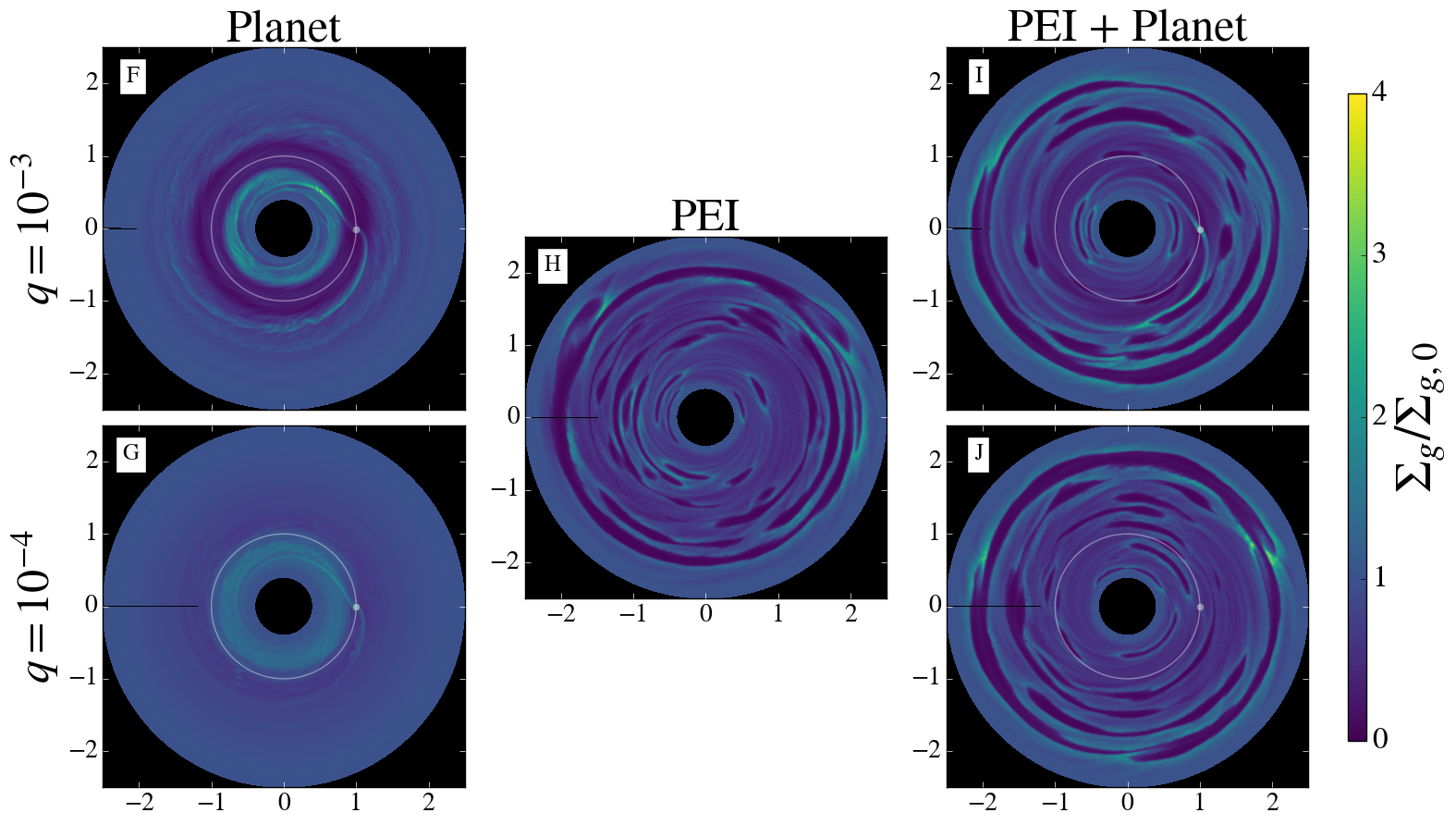}}
  \end{center}
  \caption{Dust (upper panels) and gas (lower panels) for a disk with 
    dust-to-gas ratio of unity. The disk aspect ratio is
    $h=0.05$. With the increased dust-to-gas ratio (and stronger
    effect of the drag force backreaction), the planet-only case shows increased
    turbulent-like behavior, and the wake has the appearance of
    cigarette smoke. The middle panels shows a simulation with PEI
    only. With more dust, the structures are different than in
    \fig{fig:eps01dustgas}, which is to be expected, given
    more photoelectric heating and more backreaction. 
    In the combined case (right panels), the plots suggest that
    the simulation with the higher mass planet is more depleted in
    grains than the simulation with the lower mass planet. We quantify
    this result in \fig{fig:eps1dustprofile}. In the gas
    plots (lower panels) the spiral wake of the higher mass planet is
    visible.}
  \label{fig:eps10dustgas}
\end{figure*}

\begin{figure}
  \begin{center}
    \resizebox{\columnwidth}{!}{\includegraphics{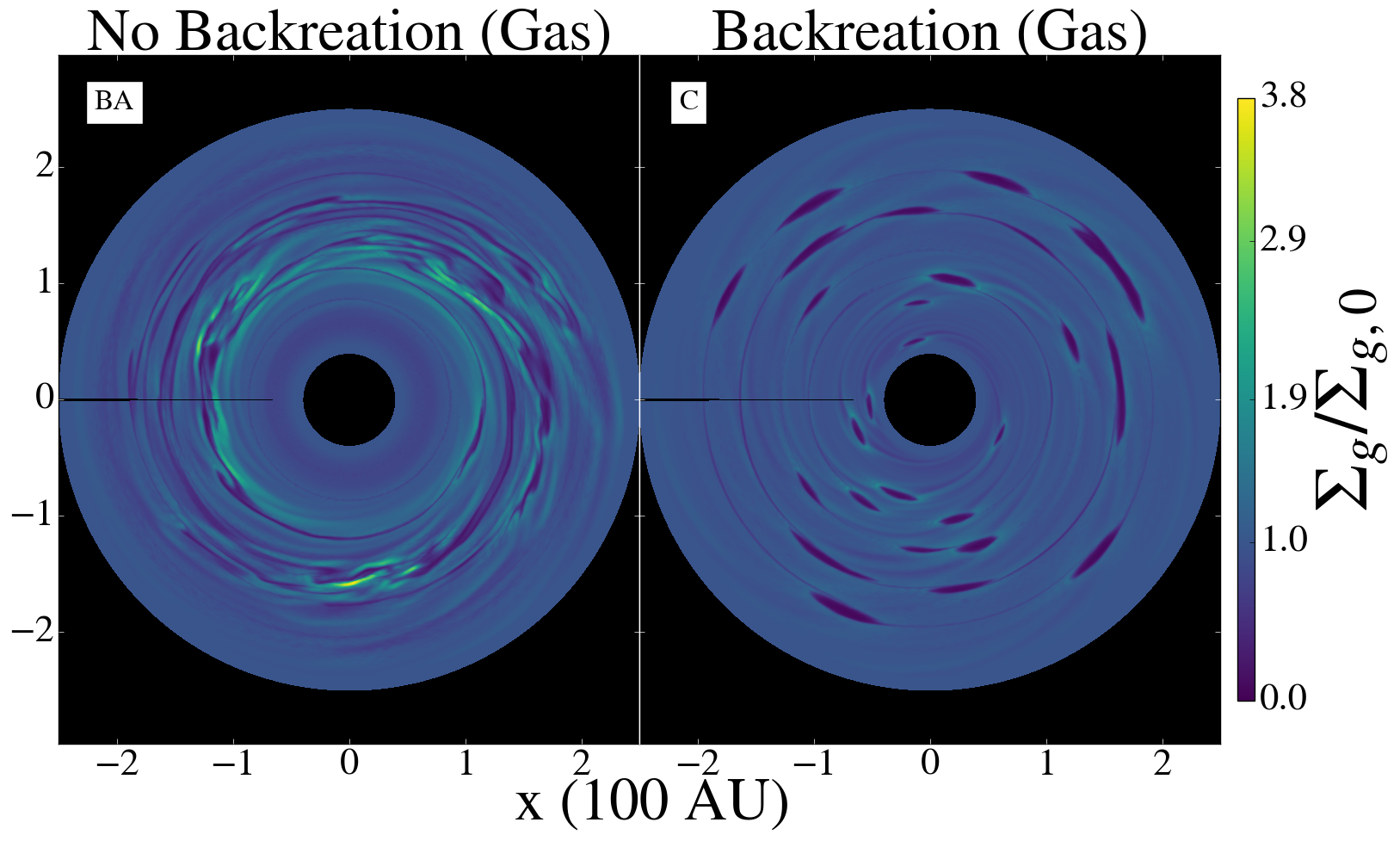}}
    \resizebox{\columnwidth}{!}{\includegraphics{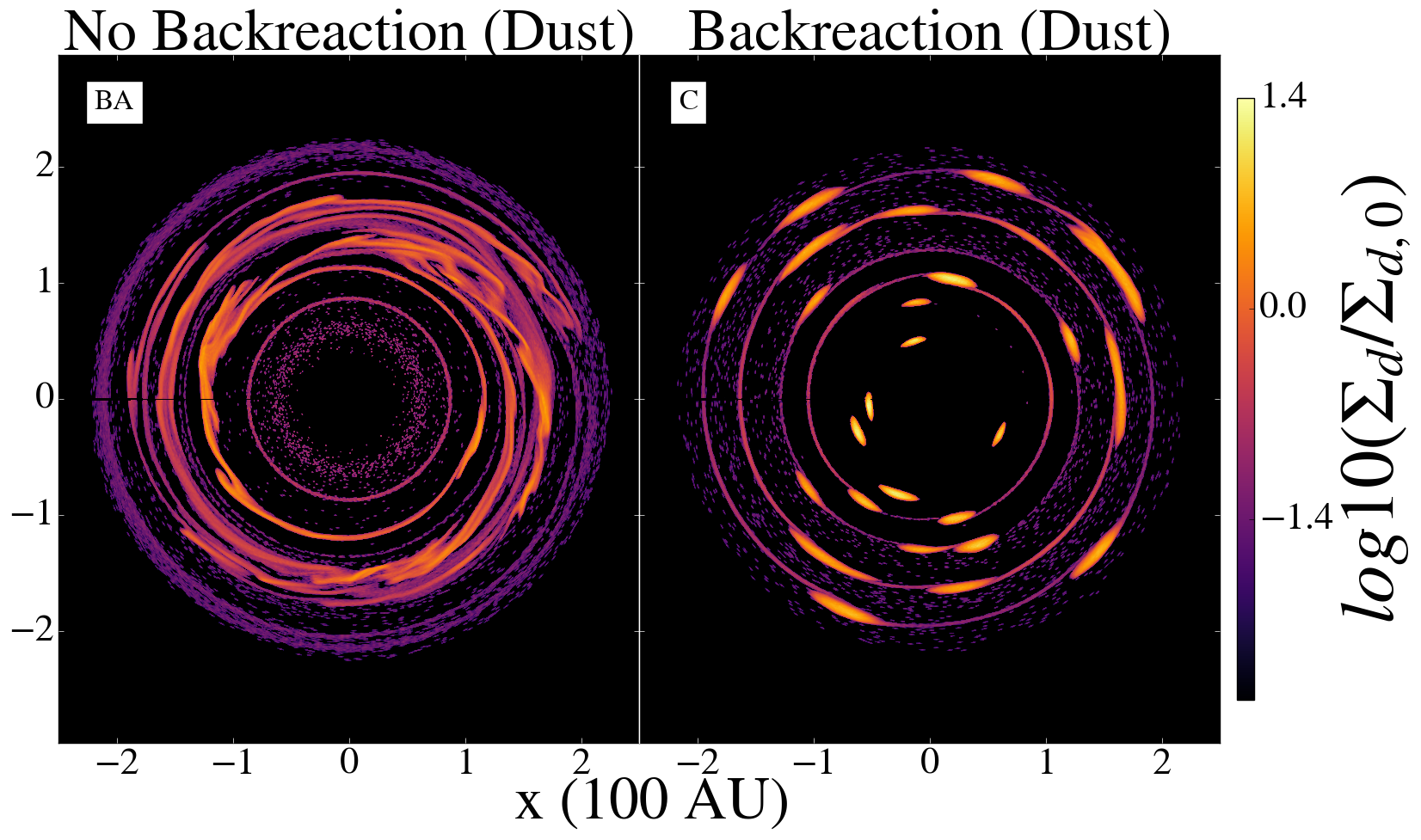}}
  \end{center}
  \caption{A comparison with backreaction active and inactive for a
    disk of $\varepsilon=10^{-1}$ and $h=0.05$. We see that the
    backreaction confines the dust into shapely and organized
    structures.}
  \label{fig:bkrk_comp}
\end{figure}

\begin{figure*}
  \begin{center}
    \resizebox{0.8\textwidth}{!}{\includegraphics{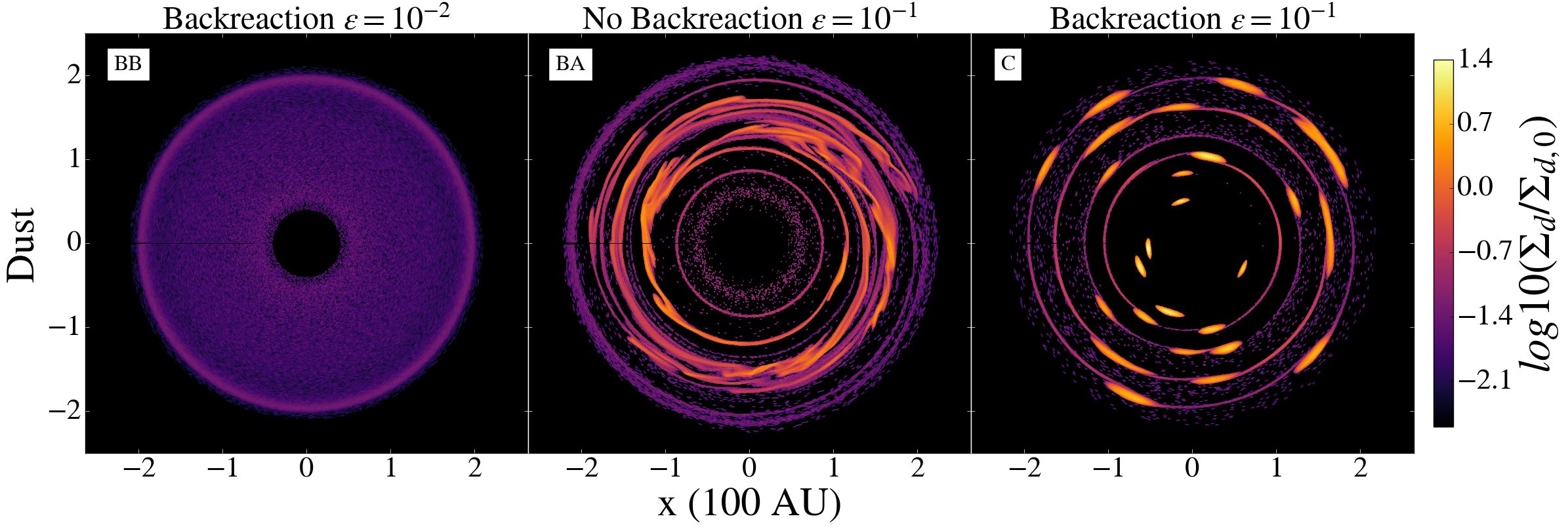}}
    \resizebox{0.8\textwidth}{!}{\includegraphics{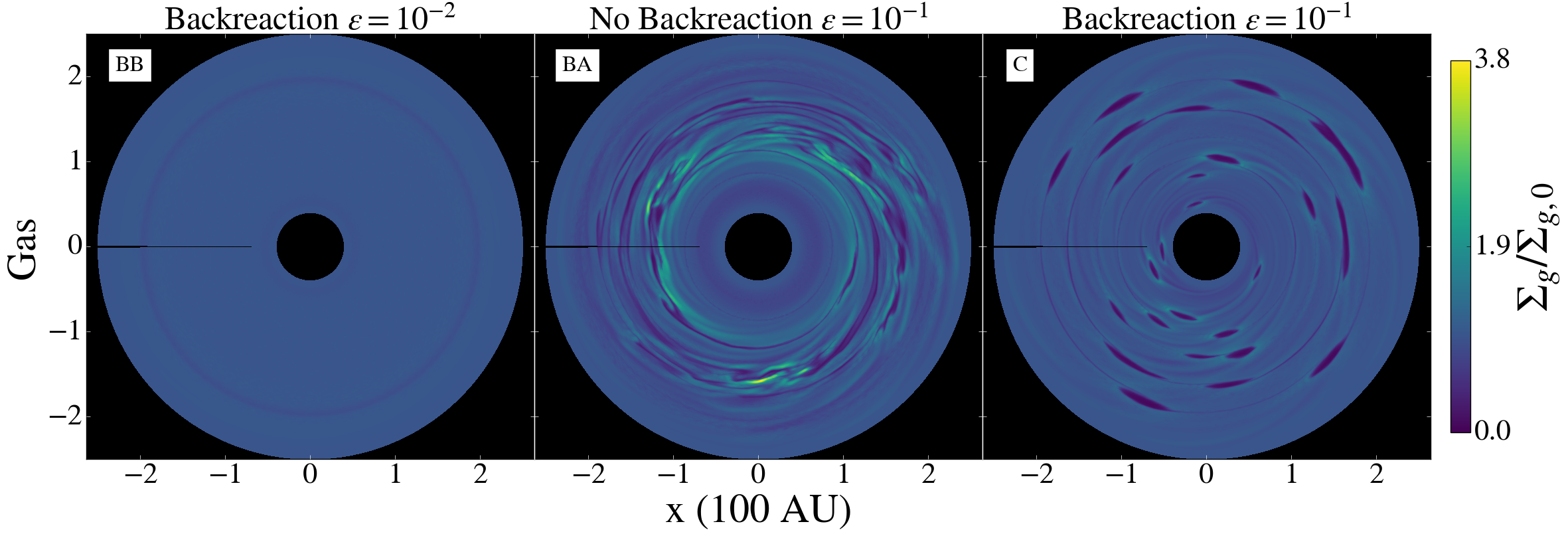}}
  \end{center}
  \caption{A comparison for the effect with and without backreaction for two dust-to-gas ratios. The two ratios are $\varepsilon=0.1$ and $10^{-2}$. The case with lower dust-to-gas ratio does not yield enough photoelectric heating to form structures, so the effect of the backreaction being turned off does not change the results.}
  \label{fig:bkrk_comp_two_epsi}
\end{figure*}

\begin{figure*}
  \begin{center}
    \resizebox{.45\textwidth}{!}{\includegraphics{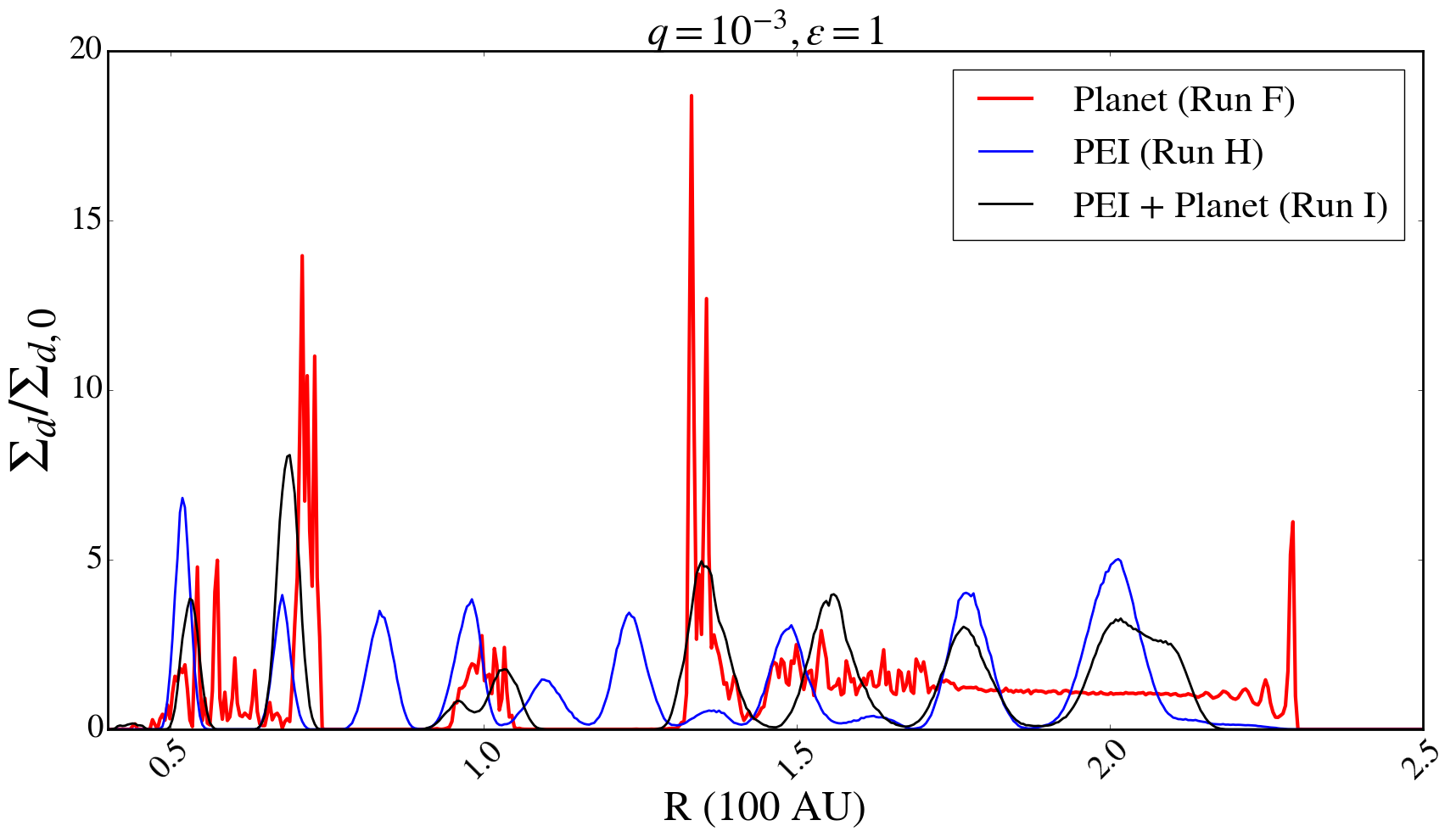}}
    \resizebox{.45\textwidth}{!}{\includegraphics{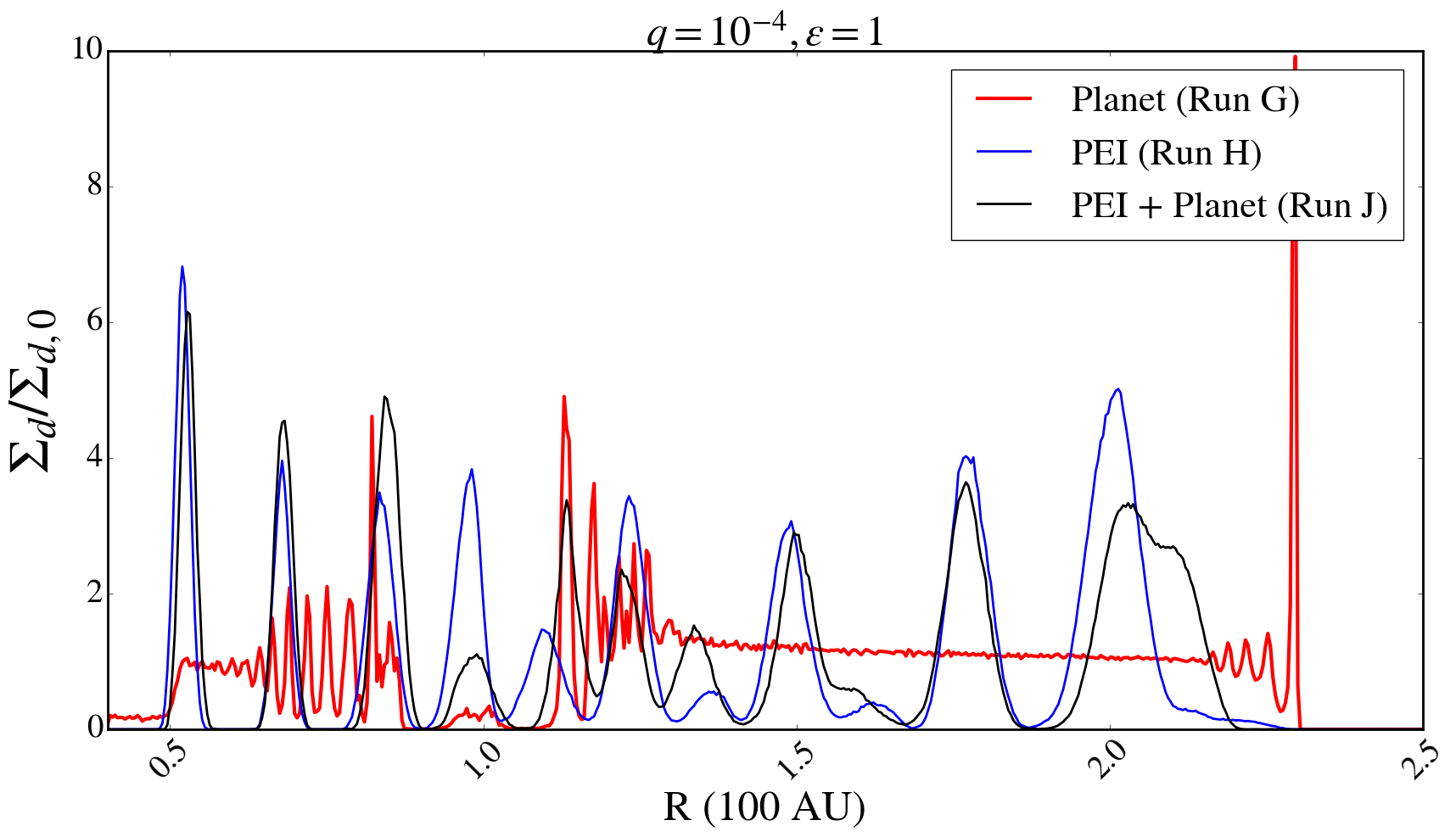}}
  \end{center}
  \caption{Same as \fig{fig:dustprofile} but for $\varepsilon=1$. The
    simulations with the Jupiter-mass planet case shows a
    discernable gap that breaks the repeating pattern of the
    photoelectric instability. {\it This is the key to distinguishing
      a planet in a disk with photoelectric instability.} The gap size
    $\delta r_{\rm gap}$ must be larger than the wavelength of the
    instability $\lambda_{\rm PEI}$. For the Neptune case, the planet
    carves a much narrower gap, which almost coincides with the 
    periodicity of the instability, making the identification of a
    planet more challenging.}
  \label{fig:eps1dustprofile}
\end{figure*}

\begin{figure*}
  \begin{center}
    \resizebox{.8\textwidth}{!}{\includegraphics{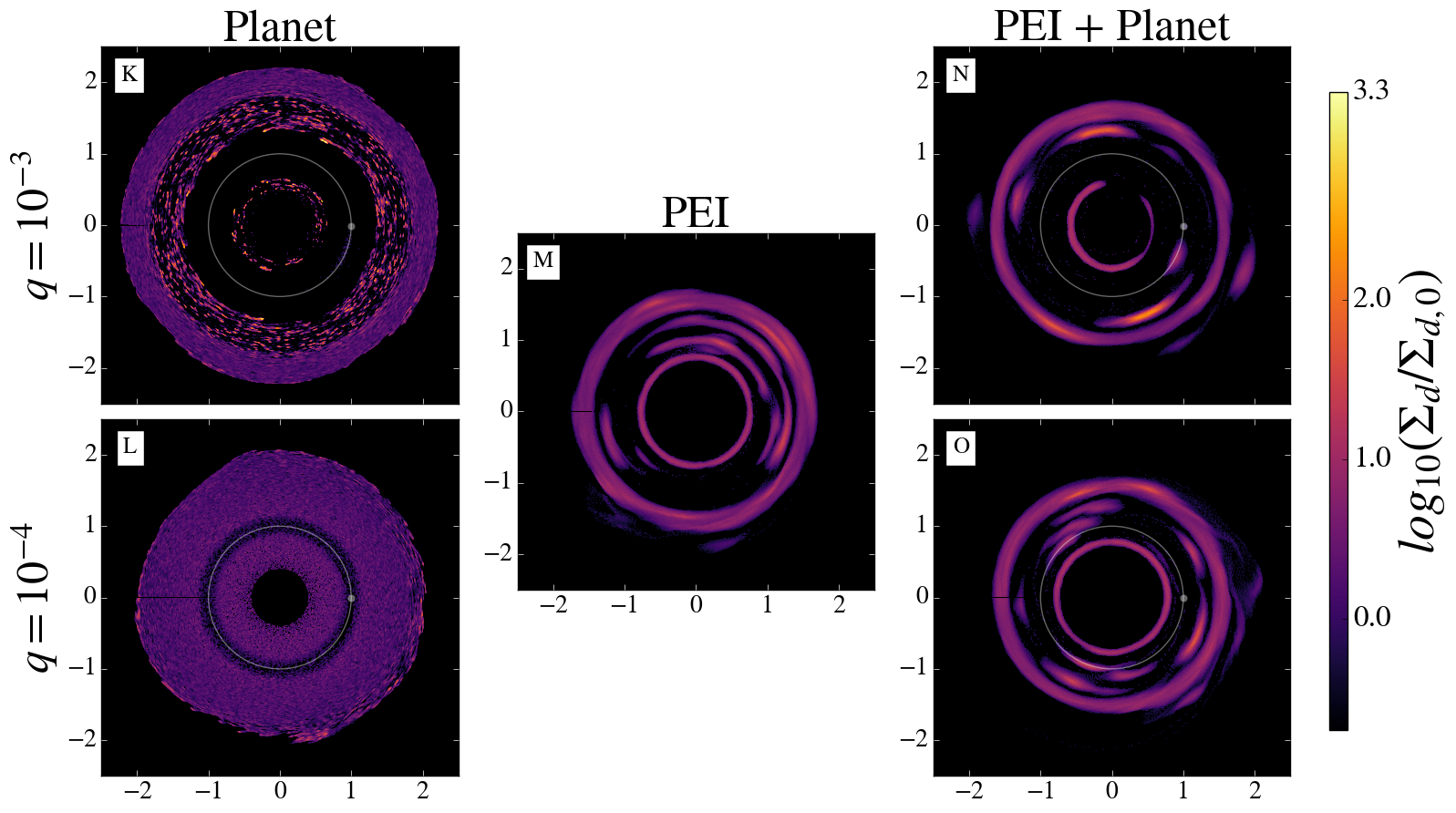}}
    \resizebox{.8\textwidth}{!}{\includegraphics{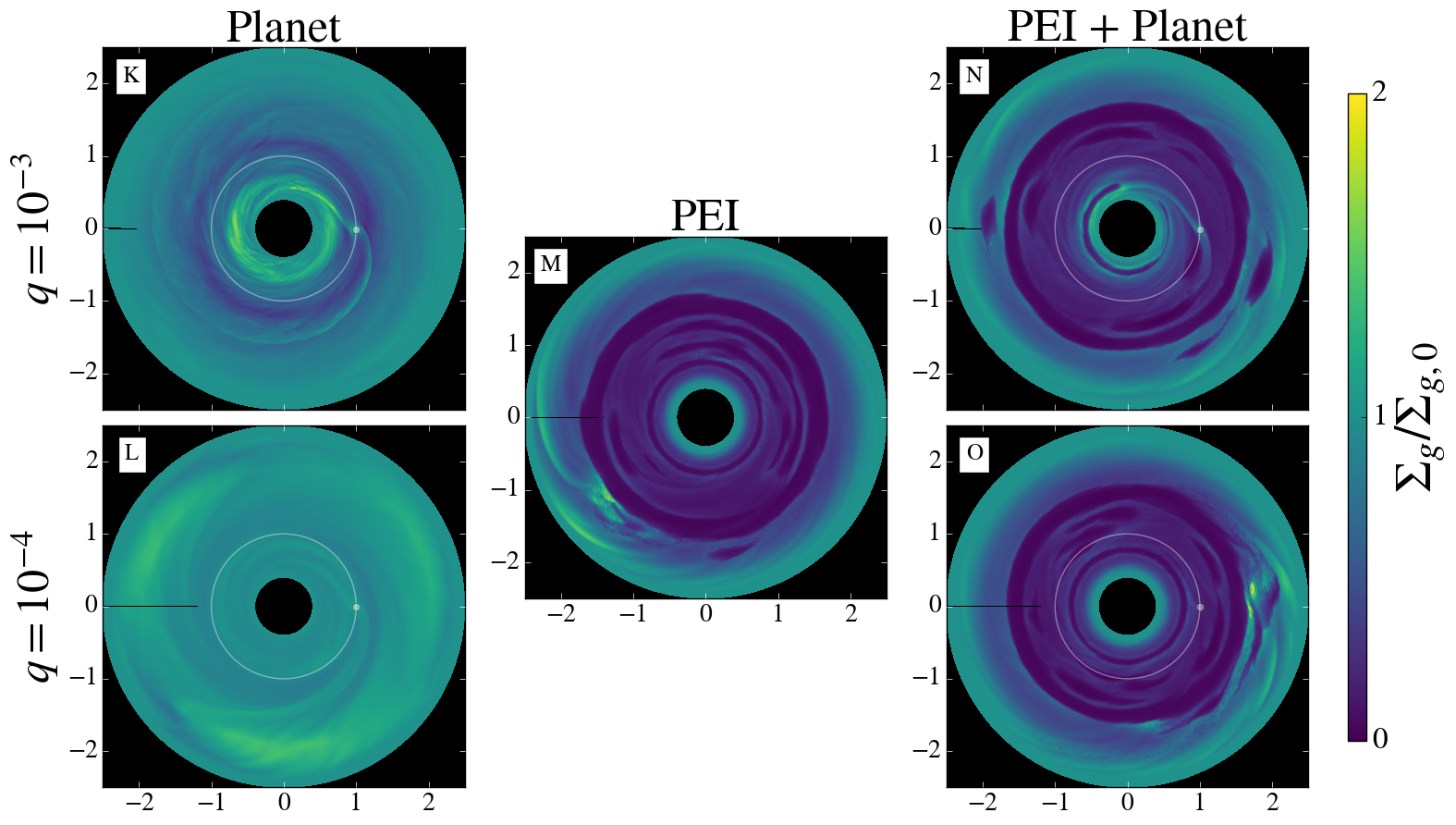}}
  \end{center}
  \caption{Same as \fig{fig:eps10dustgas}, but for $h=0.1$. For the dust
    plots we see the typical clearing out of dust, with the
    Jupiter mass carving a larger gap. The instability case now shows
    only a few arcs and one large diameter ring on the outer edges of
    the disk. The combined case shows the large disk at the outer
    boundaries with some added features. Again the Jupiter-mass planet 
    pushes grains away from its gap; and the Neptune-mass planet does
    not generate structures easily distinguishable from the PEI
    alone. With this higher temperature, the Neptune-mass planet does
    not carve a partial gap in the gas, as expected from its lower thermal mass. 
    Run L shows an asymmetry in the outer edge of the dust distribution.
    The gas plot reveals that these asymmetries are caused by vortices, in a 
    prominent $m=3$ mode. These vortices were generated by Rossby wave instability, 
    effected by the backreaction of the dragforce as the dust front drifts inwards.}
  \label{fig:h01dustgas}
\end{figure*}
\begin{figure*}
  \begin{center}
    \resizebox{.45\textwidth}{!}{\includegraphics{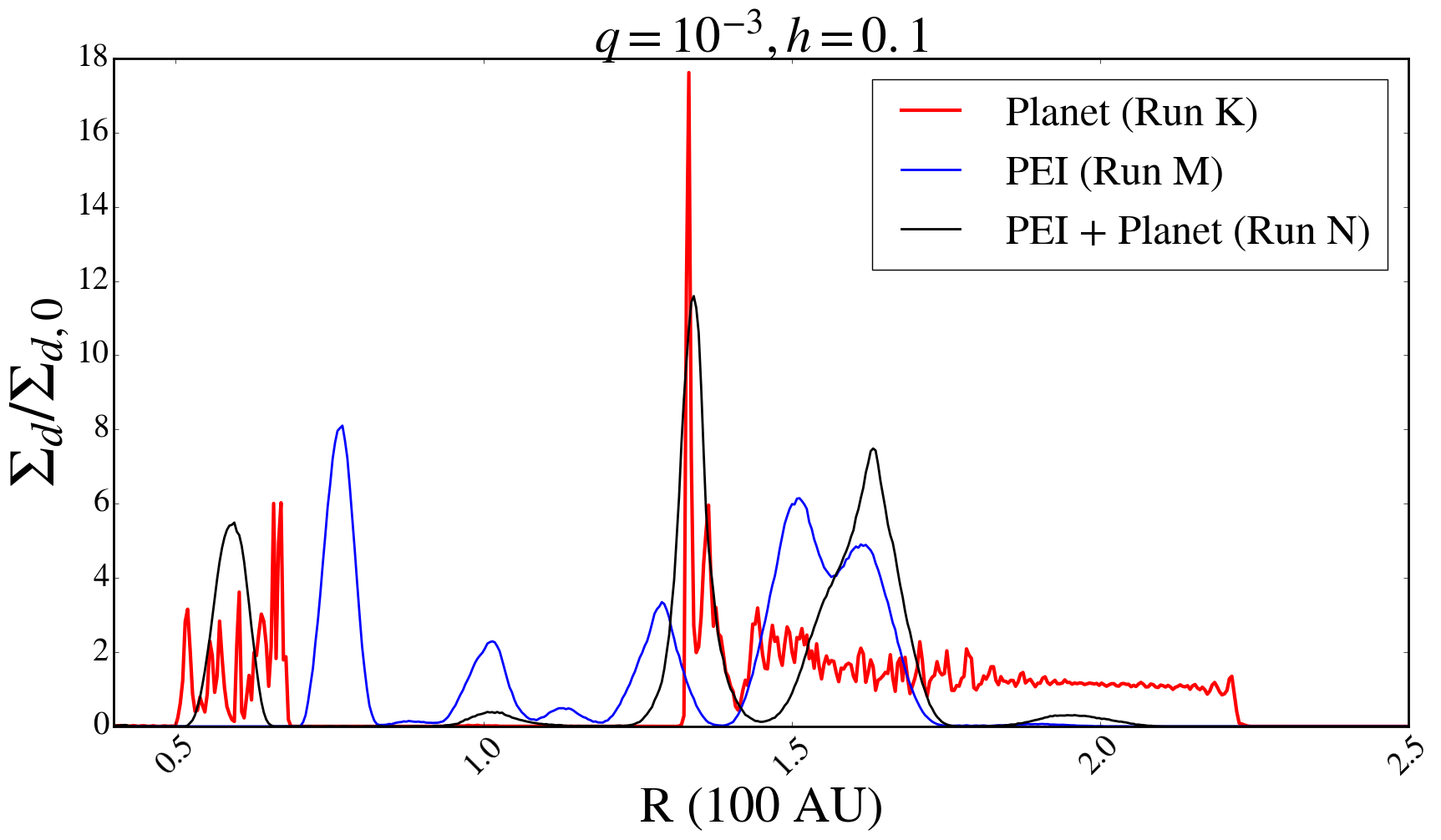}}
    \resizebox{.45\textwidth}{!}{\includegraphics{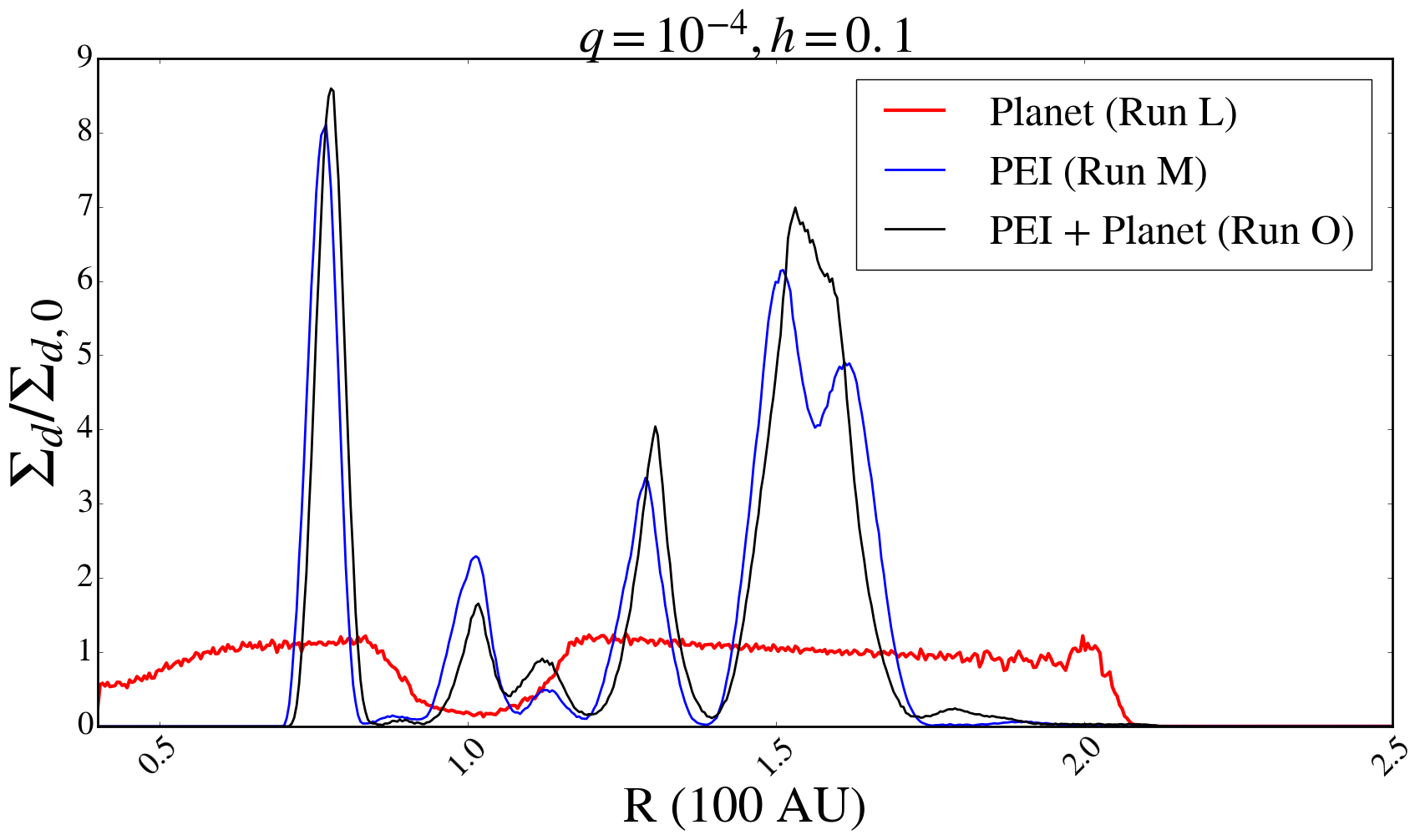}}
  \end{center}
  \caption{Same as \fig{fig:eps1dustprofile} but for $h=0.1$. Again, the
    Jupiter-mass planet shows its influence by carving a gap that interrupts the periodicity
    of the photoelectric instability. The Neptune mass planet has
    $\delta r_{\rm gap}  < \lambda_{\rm PEI}$ and is not easily
    distinguishable from the instability.}
  \label{fig:fig_1drad_avg_h01}
\end{figure*}

\begin{figure*}
 \begin{center}
   \resizebox{.8\textwidth}{!}{\includegraphics{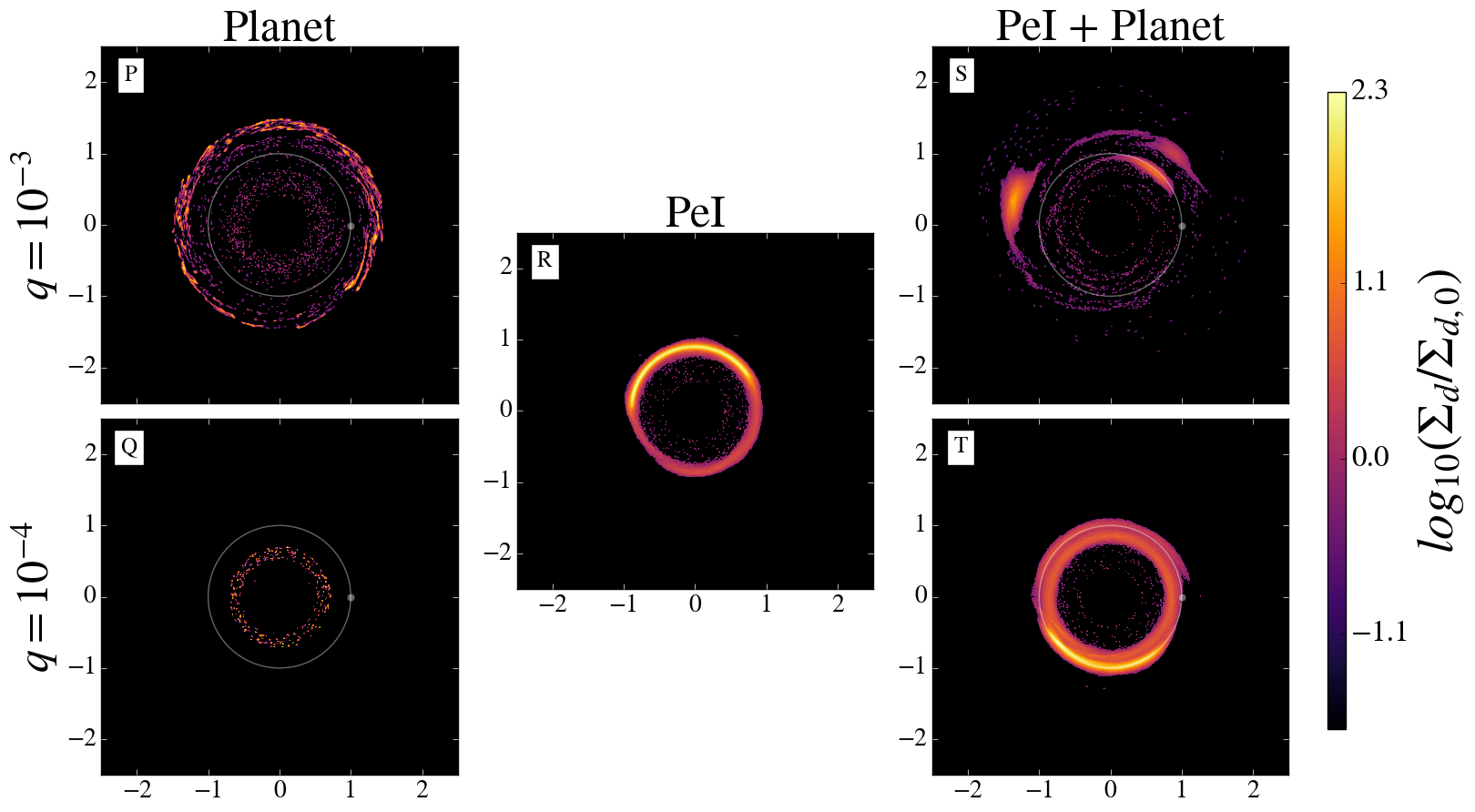}}
   \resizebox{.8\textwidth}{!}{\includegraphics{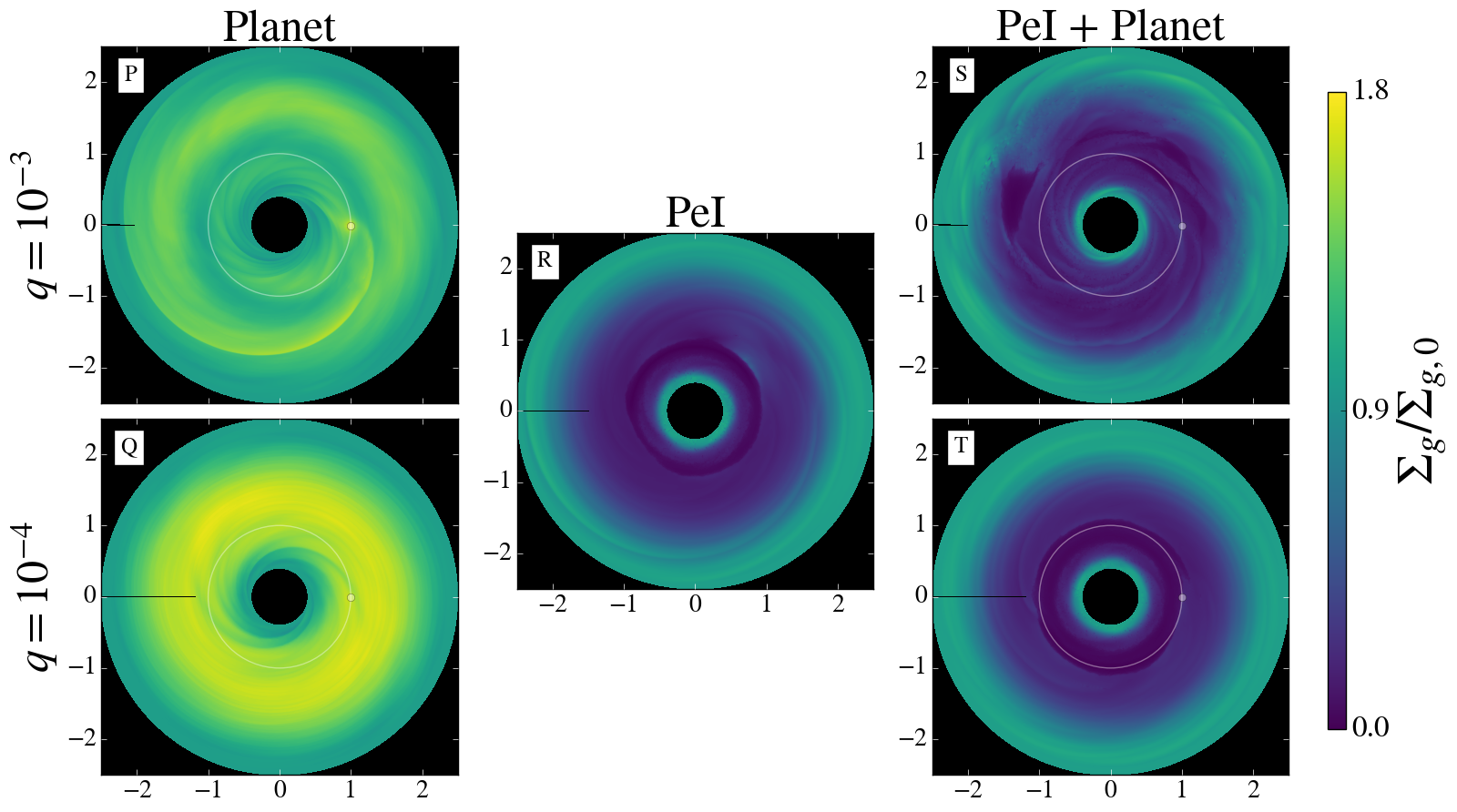}}
 \end{center}
 \caption{Similar to \fig{fig:eps10dustgas}, but for $h=0.2$. In this hot disk, neither of the planets is massive enough
     to carve a deep gas gap. For the Jupiter case a gap is carved in the dust, but for the Neptune case
     the dust streams past the planet's orbit (left plots). This is because in this
     hot disk the drift time past the planet's corotational region 
     is faster than a planetary orbit. For the Neptune case, these times
     are comparable (see \sect{sect:eps1h02}) and the dust can stream
     past the planet. The Jupiter-mass planet, with a larger
     corotational region, will be able to scatter the grains and carve
     a deep dust gap. The combined effect of a planet and the instability is barely
     distinguishable from purely PEI in this hotter disk, mostly
     because a gap is not carved and most of the dust has rapidly
     drifted inwards.}
 \label{fig:h02dustgas}
\end{figure*}

We first show the reproduction of the benchmarks of \cite{deValBorro+06} for embedded planets
of mass ratio $q=10^{-3}$ (Jupiter) and $q=10^{-4}$ (Neptune) in a disk with no dust. In the upper plots of \fig{fig:devalborro}, we plot
the gas densities. The Jupiter-mass planet carves out a deep gap in its orbit. Strong vortices are also seen
in the inner and outer edges of the gap, spawned by the Rossby wave instability \citep{Lovelace+99,deValBorro+07}.
The Neptune-mass case has the same qualitative structures, but much weaker in intensity, and the gap is shallower and narrower.
In the bottom plots of \fig{fig:qe-3gasd2g}, we plot the density divided by the initial density as a function of distance from the star. The usual gap shapes 
are reproduced. The runs with dust and photoelectric instability will be compared against these benchmark gap profiles.

%
%
\subsection{The effect of dust}
In order to study the effects of the photoelectric instability in
planet--disk interactions we need to first investigate how dust
changes the shape of the gap. The vast majority of works on
planet--disk interaction assumed either dust-free or gas-free scenarios.
Even works with both gas dust explored mainly the regime of low dust-to-gas
ratio, appropriate for primordial disks, but where the dust is mostly
passive. The regime appropriate for late stage transition disks, or
debris disk with gas, of dust-to-gas ratio near unity, has been
severely unexplored. Here we seek to find at what point does the
dust backreaction generate a noticeable effect on the gap carving
process. We explore the parameter space of dust-to-gas ratios
$\varepsilon=\rho_d/\rho_g$ and planet-to-star mass ratios
$q=M_p/M_\star$. \Table{table:allmodels} shows the explored values.\par

\subsubsection{Dust flattens the gap pressure bump}

In \fig{fig:qe-3gasd2g}, in the upper panel, we plot the gas density over
the initial gas density as a function of distance from the star. The
curves correspond to dust-to-gas ratios of 0 (gas-only), $10^{-4}$,
$10^{-3}$, $10^{-2}$, $10^{-1}$, and 1, for $h=0.05$. We see that
up to $\varepsilon=10^{-2}$ the shape of the gap does not deviate much
from the gas-only case. However, for $\varepsilon=10^{-1}$ and more
strongly for $\varepsilon=1$, a significant deviation is seen. This
corresponds to the regime where the dust begins having a significant
effect on the gas. In the lower panel we plot the local dust-to-gas
density. The dashed line marks $\varepsilon=1$. We see that only the
disks with dust-to-gas ratio of 1 or 0.1 cross this threshold. {\it We conclude that only when the local value of $\varepsilon$ reaches unity the shape of the gap is affected}, as one would indeed intuitively expect.

\subsubsection{Dependency on extent of mass reservoir}

Because of the global negative pressure gradient, the grains in the
outer disk drift inwards. At later times, the drift is also affected
by the pressure gradient generated by the planetary gap. Notice the
strong dependency of the drift speed on the
dust-to-gas ratio, with the particles started at $\varepsilon=1$
drifting slowest as they strongly affect the gas.

An immediate conclusion is that if the drift continues, and all grains
pile up at the gap edge, {\it any initial dust-to-gas ratio will eventually
achieve $\varepsilon=1$}. Indeed the simulation that started with a
global value of $\varepsilon=10^{-2}$ (yellow line) is close to this
threshold. If they can reach this value based on drift alone is
determined by the extent of the disk, i.e., by the extent of the
mass reservoir. To estimate the disk size needed to achieve $\varepsilon=1$
at the gap, let us equate the disk mass in grains beyond the gap with
the trapped mass. The former is

\beq
M_{\rm disk} = 2\pi\varepsilon_0 \varSigma_0 r_0^p \int_{r_{\rm gap}}^{r_{\rm disk}} r^{1-p} dr
\eeq

\noindent where $p \geq 0$ is the slope of the column density gradient,
$\varSigma  = \varSigma_0 (r_0/r)^{p}$. If $p$=0 then

\beq
M_{\rm disk} = \varepsilon_0 \varSigma_0 \pi \left(r^2_{\rm disk} - r^2_{\rm gap}\right).
\eeq

\noindent if we consider the gap trap to be of width $\Delta r$, then the
trapped mass is

\beq
M_{\rm gap} = \varepsilon_{\rm gap} \Sigma_0 2\pi r_{\rm gap} \Delta
r.
\eeq

Now we equate $M_{\rm disk}$ and $M_{\rm gap}$ and solve for $r_{\rm disk}$

\beq
r_{\rm disk}  = \left(\frac{\varepsilon_{\rm gap}}{\varepsilon_0}  2 r_{\rm gap} \Delta r  + r^2_{\rm  gap}\right)^{1/2}
\eeq

\noindent requiring $\varepsilon_{\rm gap} =1$, and for
${\varepsilon_0}=10^{-2}$, $r_{\rm gap}=1.4$, and $\Delta r = H$,
then $r_{\rm disk} \sim 5.5$, i.e., over 500\,AU. A steep slope will
make the demand of a larger disk even more pronounced, as the mass per ring would be less
than in the idealized case we considered.

\subsubsection{Differences in the inner disk for different planet masses and dust-to-gas ratio}
\label{sect:innerdisk}

Another feature of \fig{fig:devalborro} and \fig{fig:qe-3gasd2g} is the reversal 
seen in the behavior of the inner disk with dust-to-gas ratio for the two 
planetary masses. In the outer disk both planets show a depression in the 
gap edge as the dust-to-gas ratio increases. However, for the inner disk, while the Jupiter-mass 
planet presents the same behavior, the Neptune-mass planet does the opposite, the inner gap wall increasing 
with dust-to-gas ratio. 

The difference in behavior seems to be associated with the particle
drift. This effect is seen
in \fig{fig:drift}. The red lines show runs with $\varepsilon=1$, and
black lines show runs with $\varepsilon=0.1$. Solid lines have $b=1$ ($b$ is
the power law of the temperature, giving a global negative pressure
gradient); the dashed lines correspond to $b=0$ and thus no global
negative pressure. Without the drift the situations are similar, with
$\varepsilon=1$ (red dashed line) slightly more depressed than
$\varepsilon=0.1$ (black dashed line), because of the flattening
effect already discussed. For the runs with the global drift, a
pronounced difference is seen. As particle drift inward, to conserve angular momentum they
push the gas outwards. For higher dust-to-gas ratio (solid red line),
more gas is pushed outwards than for lower dust-to-gas ratio (black
solid line). This effect is not seen in the Jupiter-mass case because
in this case the planet carves a deep gap early enough. In the Neptune
case the shallow gap does not lead to a strong enough pressure gradient to
virtually dominate the particle motion as it does in the Jupiter case.

In \fig{fig:epsilongap} we show 2D plots of the gas density for the planet
masses $q=10^{-3}$  (upper panels) and $q=10^{-4}$ (lower
panels). Dust-to-gas ratios  $\varepsilon=10^{-2}$, $\varepsilon=10^{-1}$, and $\varepsilon=1$ 
are shown in the left and right panels, respectively. The effect of a non-zero
dust-to-gas ratio is barely distinguishable for $\varepsilon=10^{-2}$
and $q=10^{-4}$ (compare to \fig{fig:devalborro}). For dust-to-gas
ratio 0.1, the gas in the disk becomes more
turbulent at edges of the planet's orbit. The spirals that are caused
by the planet in the dustless case are smooth and laminar, whereas 
the addition of dust to the necessary threshold causes them to become
visibly turbulent at the $\varepsilon=0.1$ case. For $\varepsilon=1$ the wake and gap
look even more turbulent, with the spiral clearly passing from laminar to turbulent
at some distance from the planet, in a pattern resembling cigarette smoke.

We stress that although we consider debris disks and $\mu m$ grains, this particular result depends only on the Stokes number and dust-to-gas ratio. It is therefore applicable to primordial protoplanetary disks if pebbles and boulders ever reach these high values of $\varepsilon$.

%

%
\subsection{Disk with $\varepsilon=10^{-1}$ and $h=0.05$}

In \fig{fig:eps01dustgas}, we plot the dust (top) and gas (bottom) for a disk
with $H=0.05$ and an initial dust-to-gas ratio of
$\varepsilon=10^{-1}$. For each panel, the left columns are
simulations containing an embedded planet and no photoelectric
heating, the middle columns are simulations of a disk containing
photoelectric heating but no embedded planet,  and the right columns
disks containing both an embedded planet and photoelectric
heating. The top and bottom rows for each panel represent a
Jupiter-mass and a Neptune-mass planet, respectively. \par
The top panels show the local dust density divided by the initial dust
density, on a logarithmic scale. The cavities carved in runs A and B
manifest differently, with run A having a gap noticeably larger than
that of run B. Run A also has an overall lower dust
density throughout the disk compared to run B. These well known
results \citep{PaardekooperMellema04} are expected as a higher-mass planet
clears its orbit.
\par
Run C demonstrates the effects of photoelectric heating on a disk. The
structures created are similar to those of \cite{LyraKuchner13}, with
rings and incomplete arcs. \par
Runs D and E correspond to a disk subject to photoelectric heating and
containing a planet. Run E, which represents the Neptune analog, is
nearly identical to the control run C, containing the same structures as in the pure
instability case. We conclude that the addition of a Neptune-mass planet does not noticeably alter the shape or number
of structures from the pure instability case. Run D, which contains a
Jupiter-mass perturber, shows larger arcs and a discernable
cavity. This is another novel result. {\it When photoelectric heating is included, 
the photoelectric instability obscures structures induced by planets
unless the planet's mass is sufficiently large to carve
a noticeable gap.} The shape of the
structures caused by photoelectric instability are also changed in the
Jupiter case compared to the Neptune, likely due to the stronger
eccentricity stirring at mean-motion resonances with higher planet mass \citep{MurrayDermott99,Zhu+14}.
\par
The bottom panel shows the same plots but for the gas (gas density divided by the initial
gas density), on a linear scale. The most pronounced feature is the
anticorrelation between the dust and the gas, such that a large
concentration of dust coincides with gas depletion. Reiterating for
the gas case, we see that a Jupiter-mass perturber carves a larger gap
as well as a more prominent spiral. The spiral launched by the
Neptune-mass planet is almost lost in the midst of the structure
generated by the photoelectric instability.\par

In Figure 6, we plot an averaged linear dust density as a function of
radius from the star. The left panel coincides with a Jupiter-mass and
the right to a Neptune-mass planet. On the left panel the combined
case in solid black, creates larger concentrations of dust than the
pure instability case in solid blue. Additionally, the more massive
planet (left) carves a considerable gap at the location of the planet
(100 AU). The width of the gap is on the order of 50 AU. For the
Neptune-case the blue and black lines coincide at various locations
throughout the disk. This echoes our conclusion that a planet must
carve a substantial gap in the dust for the patterns created by the planet to be differentiated from those associated with 
photoelectric heating. The shapes shown here in the dust can be used
in tandem with actual observations in order to narrow down the
possibility of an embedded planet in a disk where the gas is subject to photoelectric heating.
%
%

\subsection{Disk with $\varepsilon=1$ and $h=0.05$}
In \fig{fig:eps10dustgas}, we plot the dust and gas for a disk with a value of
$h=0.05$ and dust-to-gas ratio of 1. For the dust, runs F and G show a gap
due to the perturber's influence, as expected. 

The increase of $\varepsilon$ to unity enhances the effect of the dust as it is easier now for dust
concentrations to affect or dominate the gas motions. The extent of
the dust distribution is larger for the $\varepsilon=1$ case than for
$\varepsilon=0.1$ because the dust dominated over the gas and stopped
drifting. The lack of drift is clear from the inner disk, where we
amount of dust in the $\varepsilon=1$ is substantial compared to the
depleted inner disk in the $\varepsilon=0.1$ case. 
Run H, which corresponds to the photoelectric instability case, shows
various arcs, rings, and gaps. This is identical to the fiducial run
of \citet{LyraKuchner13} (Figure 3 of that paper). Comparing
to the PEI case in \fig{fig:eps01dustgas}, we see that the structures are
significantly different, with mostly small-scale arcs in the
$\varepsilon=0.1$ case, and no complete rings. This is 
similar to the backreaction-free models of \citet{LyraKuchner13}, who found that
backreaction was needed to stabilize and maintain complete rings. In
that paper we examined only $\varepsilon=1$ for the 2D runs, finding
that in a run without the backreaction (Suppl. Fig 6 of that paper),
the disk broke into several clumps. We explained it in terms of the
tendency of the gas to counter-rotate in and around regions of high
pressure, which is turn is disrupted by the backreaction. For the
$\varepsilon=0.1$ case, the backreaction is weaker than in the
$\varepsilon=1$ case, so the ability of the PEI to structure the dust into
rings is diminished. Run C, however, is more self-organized than the
no-backreaction run in \citet{LyraKuchner13}, that appears strongly 
turbulent. The differences between these runs is not just the presence
of backreaction, but also the lower amount of dust, that leads to less
heating.

In \fig{fig:bkrk_comp}, we check
the effect of the dust backreaction for a simulation of
$\varepsilon=0.1$. The upper panels show the gas while the lower
panels show the dust; the left panels are runs without backreaction,
the right panels runs with backreaction. We see that the backreaction
indeed confines the dust into shapely and organized structures. The
run without backreaction shows spiral features and accumulation of
power in high azimuthal wavenumbers, albeit at a lesser degree than in
the $\varepsilon=1$ run in \citet{LyraKuchner13}. We expect that for
$\varepsilon=10^{-2}$ the backreaction will be less important and thus
the structures will be more turbulent. {\it The conclusion is that
well-organized PEI structures should be expected only in disks with
high dust-to-gas ratio, higher than primordial.}

Runs I and J show a disk with PEI and a planet, of $q=10^{-3}$ and $q=10^{-4}$,
respectively. The more massive perturber clears out a noticeable
gap in the dust. Comparing run J to run H we see no noticeable differences.
This echoes the conclusion that only planets that carve substantial
gaps can be disentangled from the effect of PEI. 
For the gas in \fig{fig:eps10dustgas}, we see similar results to \fig{fig:eps01dustgas}.
Runs I \& J, which include a planet and PEI, show one important
difference. The Lindblad spiral of the Neptune analog, which had
already become weaker in run E ($\varepsilon=0.1$), has all but
disappeared in run J ($\varepsilon=1$), washed out by the stronger
effect of the PEI. The Jupiter-mass planet in run I is still visible
in the gap showing a gap and the Lindblad spiral. 
\fig{fig:eps1dustprofile} is the same as \fig{fig:dustprofile}, but now the dust-to-gas ratio is
equal to unity. For the Jupiter mass planet the dust averages are shown on the left hand side.
The dashed red line and the solid black line both contain a planet, and show a
depletion of the dust around the location of the planet at $r=1$. The solid blue
line is for a simulation with only PEI, and dust densities are unchanged where
$r=1$.
The Neptune case is plotted on the right. Again, the red dashed line and solid
black line are for simulations with a planet. The pure instability case, in the solid blue line,
coincides almost identically to the case including a planet and
PEI, except for the peak near the planet, inside the gap. In this
simulation, a case may be made that with sufficient accuracy an
observation would be able to distinguish between a PEI (blue line) and a
planet+PEI (black line) case, judging by the lower dust peak at $r=1$ in
the planet+PEI simulation, so we do not categorically state that the
case is hopeless for these lower mass planets, only that significant
effort from the observational side would be required to detect such
subtle a difference. 
Our results help on the effort to remove the signatures from
hydrodynamical effects in order to unambiguously detect planets in
disks, and may be applicable to disks such as HD 141569A and 49 Ceti,
where the structures seen qualitatively match those produces by PEI.\par
%
%

\subsection{Dependence on thermal mass}

Because our results depends on the ability of the planet to carve a
gap, the parameter that controls the observability is not the mass of
the planet, but the thermal mass. The thermal mass is that for which
the Hill radius of the planet equals the scale height of the
disk. That yields a mass ratio 

\beq
q_{\rm th} = 3 h^3 
\eeq

For $h=0.05$, a Jupiter-mass planet is equivalent to $q=8q_{\rm th}/3\approx 2.7 q_{\rm
  th}$, and the $q=10^{-4}$ planet to $\approx 0.27q_{\rm th}$. 

Increasing the aspect ratio $h$ by a factor 2 (to $h=0.1$) increases
the thermal mass by a factor 8. The Jupiter-mass planet
should become equivalent to $q_{\rm th}/3$, roughly equivalent to the
$q=10^{-4}$ planet in the $h=0.05$ case.

We have found the condition that disentangles the carving of a gap from the
structures created from the photoelectric instability.
We take the radius of the dust gap and the frequency of the photoelectric
instability to be $r_{gap}^{(d)}$ and $\lambda_{\rm PEI}$, respectively.
In order for the two processes to be differentiated, we must have that
$r_{gap}^{(d)} > \lambda_{\rm PEI}$. In this regime, the gap is large enough
that peaks caused by the instability are within the gap, as is seen in
\fig{fig:fig_1drad_avg_h01} for the Jupiter case. Comparing this to the
Neptune case in the same figure $r_{gap}^{(d)} \approx \lambda_{\rm PEI}$,
showing no concrete difference between the PEI and combined case.

In the next subsections, 
we explore the role of temperature.

\subsubsection{Disk with $\varepsilon=1$ and $H=0.1$}
\label{sect:eps1h01}

\fig{fig:h01dustgas} shows the dust and gas for a disk of aspect ratio
$h=0.1$. The top panel plots the dust and the bottom panels the gas, 
as in the previous cases. The Jupiter mass planet has thermal mass
$0.33$, and the Neptune mass planet $0.03$. We see that indeed
the Neptune analog does not carve a gap in the gas, staying deeply
embedded. The Jupiter-mass planet carves a shallower gap, as expected
from its thermal mass now being lower than 1. 

The dust panel of run L shows departure from axisymmetry and an enhanced dust drift compared
to the Jupiter case for elsewise the same parameters (run K). Examining
the gas plots below, we notice three prominent structures in run L, at
$r=2$. These structures resemble vortices, but if so, it is curious
that they are not trapping dust. This should only happen if the
vortices are triggered after the dust has drifted inwards, so that
there is no dust to be captured. Indeed, the outer boundary of the
dust distribution, seen in the dust panel of run L, is inwards of the
radial location of the vortices as seen in the gas panel of the same
run. 

It is suspicious that this phenomenon occurs near the boundary. Yet
run K does not show the vortices, which rules out boundary conditions
as culprit. Instead, we conjecture that what we are seeing is Rossby
Wave Instability triggered by the dust drift in a high dust-to-gas
ratio. As the dust drifts inwards, the dust front constitutes a sharp
discontinuity. At high dust-to-gas ratio, the backreaction of the drag
force has significant dynamical relevance, so the gas has a different
rotational velocity immediately inwards and immediately outward of
the dust front. This difference eventually becomes sufficient to
trigger the RWI. This is a novel effect, {\it RWI triggered at a dust
front}. While in this case the dust front is artificial (it would not
exist if there was a supply of dust from the outer disk), there are
physical situations where a sharp dust discontinuity exists, as in,
e.g., the birth ring model. In disks with a birth ring and significant
gas, we thus predict that RWI should be triggered in the gas. If the
vortices disturb the population of grains in the same way, the birth
ring would turn eccentric, as the dust front in run L. 

The Jupiter mass planet does not show the same phenomenon
because of the reduced dust drift, so the front is still located at
the boundary region of the run, where the quantities are driven back
to the initial conditions. For the $h=0.05$ run, we conjecture that the
same process would be present, but the dust has not drifted far enough
inwards yet for the dust front to be resolved out of the boundary
region. The dust drift is proportional to the sub-Keplerian reduction
parameter $\eta$, which is proportional to $h^2$. Increasing h by a
factor 2 increasing the drifting time by a factor 4, making the effect
noticeable.  Also, the resolution was probably too low to resolve
the narrow band of the instability. Increasing $h$ to 0.1 doubled the
width of the transition, allowing the modes to be resolved. 

In run M, which is a simulation with PEI and no planet, we see less arcs and more
rings than in the $h=0.05$ case, the structures are thicker also with
the increased value of $H$. The gas disk is empty in the center,
owing to an increased heating, according to
\eq{eq:reftemperature}. The gas expands due to the higher temperature,
thinning out. At the boundaries it retains the high initial density
value because of our boundary condition that drives the quantities
back to their initial values. 

The combined case of a planet and PEI, shown in runs N and O, shows rings and
arcs as well. Run N shows a larger gap, with a lack of dust in the immediate
area, apart from two concentrations at 12 and 5 o'clock. 
Run O, for a lower mass planet shows various arcs 
and a non-discernable gap in the dust. 

%

%
%
%

\subsubsection{Disk with $\varepsilon=1$ and $H=0.2$}
\label{sect:eps1h02}

Finally, we consider a disk with a scale height of $h=0.2$, which is
on the hotter side for primordial disks, but fairly typical for the scale
heights of debris disks with gas, including $\beta$\,Pic. We plot in
\fig{fig:h02dustgas}, on the top panel, the dust structures, and in the
lower panel the gas structures. For the Jupiter case (upper left) 
we see a gap carved in the dust. However, for the Neptune 
case we only see a small ring inward of the planet and nothing
outwards. Clearly the dust drift is more intense for the Neptune
case. The reason for this is twofold. Firstly, as seen in the gas
plots (lower panel), the Jupiter-mass planet, though having a thermal
mass of 0.04, carves a very shallow gap. The Neptune mass planet has
thermal mass 0.004 and is unable to open a gap. The high density
outside the Jupiter-mass planet traps the drifting grains. In the
Neptune-mass planet case, they stream past the planet orbit and are trapped at the
density maximum inwards of it. 

It is curious that although there is not a gas gap, the grains are able
to stream past the planet, bypassing the dust filtering that planets 
usually do. This seems to happen because the drifting time is faster
than a planetary orbit. We can compare the time it takes a grain to
drift past the corotational region of planet to the orbital period of
the planet 

\beq
\Theta  = \frac{t_{\rm drift}}{t_{\rm orb}}
\eeq

If $\Theta > 1$, the drifting is slow and the planet is able to
scatter the particle. If on the other hand $\Theta < 1$ the planet is
too slow and the grain can drift past it. Considering the drift velocity of a grain \cite{Youdin10}

\beq
v_r = -\frac{2\St\,\eta\,v_k}{1+\St^2},
\eeq

\noindent where $v_k=\varOmega_k r$ is the Keplerian velocity and 

\beq
\eta \equiv = -\frac{1}{2\rho\varOmega^2r}\frac{\partial p}{\partial r}
\eeq

\noindent is the sub-Keplerian parameter. The time it takes to drift is $v_r/2x_s$ where $x_s$ is the
half-width of the corotational region, given  by \citet{PaardekooperPapaloizu09}

\beq
x_s^2 = 1.68 r_p^2 \frac{q}{h} 
\eeq

\noindent and $r_p$ the planet's location. Solving for $\Theta$, 

\beqn
\Theta &=& \frac{\varOmega  x_s}{\pi v_r }\\
&\approx&0.4 \ q^{1/2} h^{-5/2} \chi^{-1} \St^{-1} \left(1+\St^2\right) 
\eeqn

\noindent where $\chi=\alpha+\beta$ with $\alpha=-\partial\ln \rho/\partial\ln r$ and
$\beta=-\partial\ln T/\partial\ln r$.  For $q=10^{-4}$, $h=0.2$, $\St=1$
and considering $\alpha+\beta=1$, we get $\Theta=0.5$. For the
Jupiter-mass planet, $\Theta=1.4$. Thus, it is reasonable to expect that
in the Neptune-mass case the $\St=1$ grains can drift past the corotational
region in this hot disk but that the Jupiter-mass planet, with a
larger corotational region, will be able to scatter the grains and carve a
deep dust gap {\footnote{ Built-in in this estimate is our
    assumption that the Stokes number is constant. In fact, if a
    planet carves a deep gas gap, the Stokes number should go to
    infinity (as there is not gas, there is no drag). As such, this
    analysis applies only to partially depleted gaps (low and
    intermediate-mass planets).}}. In the simulation with photoelectric instability (run R), we see a
much denser ring of dust in the center of the domain. The combined case
of instability plus planet (runs S and T)  are shown in the plots in
the right hand side. For the massive planet (run S) the main differences are
that the grains, instead of being scattered roughly uniformly as in
run P, now concentrates in arcs. In addition, an arc coinciding with
the location of the Lagrange point L5 is prominent. The lower mass
planet (run T) seems indistinguishable  from the pure instability case
(run R). 

The behavior of the gas (lower plots) resembles the situation with
$h=0.1$. The gas density is much lowered throughout most of the disk
due to the enhanced photoelectric heating, being driven back to the
initial value at the boundaries. A spiral is barely discernible for
Jupiter-mass planet. No clear distinction is present for the lower
mass planet. 

\section{Caveats}

The presented models are admittedly simplified. In this section, we
state what we consider the main limitations of our calculations to
be.

The most stringent limitation of the models is the 2D
approximation. The twodimensionality imposes that the vertical
direction is degenerate, so we do not resolve
the vertical scale height. As a result the backreaction of the drag
force is largely underestimated. In 3D, the midplane grain layer would
be far denser due to sedimentation. 

Another caveat is that we kept the Stokes number constant. In reality,
the Stokes number in the Epstein regime should depend on density and
temperature, according to \eq{eq:tauf-omega}. The effect of a
self-consistent Stokes number in the PEI has not been explored, but it
is not expected to be significant -- as dust concentrates they
increase the gas temperature, which also decreases
the density through gas expansion. The former
decreases $\St$, the latter increases $\St$, so both combined mitigate
the effect on $\St$.

%
%
\section{Conclusions}
\label{sect:conclusions}

In this paper we have considered for the first time disk-planet
interaction in the presence of photoelectric instability. We also study for the first time planet--disk interaction for
disks of dust-to-gas ratio around unity, where gas and dust have
similar inertia and none dominate over the other. 

We find that the gaps carved in disks with dust and gas deviate from the
  gas-only case when the local dust-to-gas ratio approaches unity. 
  The gap walls, because they are dust traps, are the places where 
  the deviation occurs. At global dust-to-gas ratio $\varepsilon=10^{-2}$, the effect starts to become
  noticeable as a slightly reduction in the height of
  the gap wall. For $\varepsilon=10^{-1}$ and $\varepsilon=1$, the gap
  wall is significantly depressed as the local dust enhancement goes
  well above unity. This result does not depend on photoelectric
  instability and thus applies to optically thick (primordial and gas-rich
  transition) disks as well.  

Our simulations show that when photoelectric instability is
included, the features produced by low-mass planets are not
conspicuous enough to be distinguishable from the features produced by
the instability alone. The signature of gap carving and a conspicuous
spiral are the tell-tale signatures of planets, as
is usual for conventional transition disks. The most clear sign of a planet is a
radial clearing in the PEI-induced structures, wider than
the typical separation between them. That is, we can distinguish a
planet amid of photoelectric instability structures if the planet carves a
dust gap larger than the wavelength of the photoelectric instability. Quantitatively, we
can state that the criterion to distinguish a planet in a disk with
photoelectric instability is $\Delta r^{(d)}_{\rm gap} > \lambda_{\rm PEI}$, where $\Delta r^{(d)}_{\rm
  gap}$ is the radial width of the dust gap and $\lambda_{\rm PEI}$ the
wavelength of the instability. The quantity $\Delta r^{(d)}_{\rm gap}$
is the simpler one to quantify. The boundary of the dust gap is located at the 
edge of the gas gap, that functions as a dust trap. As for
$\lambda_{\rm PEI}$, linear instability analysis predicts growth even
for $k\rightarrow\infty$ \citep{LyraKuchner13}. Regularizing with viscosity produces the most
unstable wavelength of the order of the scale of viscous dissipation,
which is very small in primordial and transitional disks, but of

order AU inlate transition and debris disks. Yet, a 3D run in \citep{LyraKuchner13}
shows ring spacing of $\approx 0.1H$ for a simulation with grid
resolution $dx=H/256 \approx 0.004$, i.e., $\lambda_{\rm PEI}/dx \approx 25$,
and thus far from the viscous range. A specific prediction for
$\lambda_{\rm PEI}$ is still lacking in the literature, a situation
that is compounded by the fact that the PEI also exists in the
nonlinear regime.

Our simulations also show that the PEI leads to well-defined sharp
structures only for high dust-to-gas ratios, because of the pivotal
role of the drag force backreaction in confining the
dust. Simulations without backreaction, or with too little dust, 
$\varepsilon \approx 10^{-1}$, easily break into spiral
structures and turbulence. We conclude that well-organized PEI 
structures should be expected only in disks with
dust-to-gas ratio higher than primordial. 

Finally, incomplete arcs are not produced
in the simulations with planets alone: they require the photoelectric
instability. However, we did not consider situations with self-gravity,
which is thought to be a component in a possible mechanism to maintain incomplete arcs in
the ring system of Neptune \citep{NamouniPorco02,Tsui18}, albeit
unlikely in the case of debris disks, of very low mass, to produce a
noticeable effect.

The models of \citet{Richert+18} show that radiation pressure can have a significant effect on the global dust distributions of optically thin disks. The models we present in Section~\ref{sect:results} are most straightforwardly applicable to optically thin disks around low-mass stars (spectral types K and M) and white dwarfs, whose low luminosities imply a minimal amount of radiation pressure even on small, well-coupled ($\tauf\varOmega=1$) grains. On the other hand, for more luminous stars, the greater radiation pressure will likely cause the disk morphologies to deviate substantially from the models presented in this paper. Nonetheless, our models and those of \citet{Richert+18} demonstrate that radiation pressure, massive planets, and the photoelectric instability can all play an important role in shaping optically thin disks, suggesting that future models should more extensively explore the combination of these and other effects. In particular, the MRI is another potentially important process in such systems \citep{KralLatter16} whose interplay with these other processes remains poorly explored.

%
\acknowledgments
W.L. acknowledges support of Space Telescope Science Institute through
grant HST-AR-14572, and the NASA Exoplanet Research Program through grant
16-XRP16\_2-0065. M.K. acknowledges
support provided by NASA through a grant from
the Space Telescope Science Institute (HST Cycle 21 AR13257.01),
which is operated by the Association of Universities
for Research in Astronomy, Inc., under NASA contract
NAS 5-26555. The simulations presented in this paper utilized the
Stampede2 cluster of the Texas Advanced Computing Center (TACC) at The
University of Texas at Austin, through XSEDE grant TG-AS140014. We are indebted to the referee,
Dr S\'ebastien Charnoz, for thoughtful
  comments that helped improve the manuscript.
%

  
\bibliographystyle{apj}
\bibliography{unitypaper}

\end{document}